\documentclass[11pt, notitlepage]{article}

\usepackage{todonotes}
\usepackage{a4}

\usepackage{xcolor}
\usepackage{algorithm}
\usepackage[noend]{algpseudocode}

\definecolor{TUMblue}{RGB}{0, 101, 189}
\definecolor{TUMlightblue}{RGB}{100,160,200}
\definecolor{TUMgreen}{RGB}{162,173,0}
\definecolor{TUMorange}{RGB}{227,114,034}
\definecolor{TUMivory}{RGB}{218,215,203}

\usepackage{natbib}
\PassOptionsToPackage{hyphens}{url}\usepackage{hyperref}

\usepackage{hyperref}
\hypersetup{
colorlinks=true,
linkcolor=TUMblue,
citecolor=TUMblue,
filecolor=TUMblue,
urlcolor=TUMblue
}

\usepackage{etoolbox}
\makeatletter

\pretocmd{\NAT@citex}{%
\let\NAT@hyper@\NAT@hyper@citex
\def\NAT@postnote{#2}%
\setcounter{NAT@total@cites}{0}%
\setcounter{NAT@count@cites}{0}%
\forcsvlist{\stepcounter{NAT@total@cites}\@gobble}{#3}}{}{}
\newcounter{NAT@total@cites}
\newcounter{NAT@count@cites}
\def\NAT@postnote{}

\def\NAT@hyper@citex#1{%
\stepcounter{NAT@count@cites}%
\hyper@natlinkstart{\@citeb\@extra@b@citeb}#1%
\ifnumequal{\value{NAT@count@cites}}{\value{NAT@total@cites}}
{\ifNAT@swa\else\if*\NAT@postnote*\else%
\NAT@cmt\NAT@postnote\global\def\NAT@postnote{}\fi\fi}{}%
\ifNAT@swa\else\if\relax\NAT@date\relax
\else\NAT@@close\global\let\NAT@nm\@empty\fi\fi% avoid compact citations
\hyper@natlinkend}
\renewcommand\hyper@natlinkbreak[2]{#1}

\patchcmd{\NAT@citex}
{\ifNAT@swa\else\if*#2*\else\NAT@cmt#2\fi
\if\relax\NAT@date\relax\else\NAT@@close\fi\fi}{}{}{}
\patchcmd{\NAT@citex}
{\if\relax\NAT@date\relax\NAT@def@citea\else\NAT@def@citea@close\fi}
{\if\relax\NAT@date\relax\NAT@def@citea\else\NAT@def@citea@space\fi}{}{}

\makeatother

\usepackage{amssymb, graphicx}
\usepackage{subfigure}
\usepackage{amsmath}
\usepackage{amsthm}
\usepackage{mathabx}
\usepackage{verbatim}
\usepackage{bbm}
\usepackage{dsfont}
\usepackage{geometry}
\usepackage{pdflscape}
\usepackage{multirow}
\usepackage{aliascnt}

\newcommand{\mynewtheorem}[2]{
\newaliascnt{#1}{dummy}
\newtheorem{#1}[#1]{#2}
\aliascntresetthe{#1}
\expandafter\def\csname #1autorefname\endcsname{#2}
}

\theoremstyle{definition}
\mynewtheorem{thm}{Theorem}
\mynewtheorem{defi}{Definition}%[section]
\mynewtheorem{lem}{Lemma}%[section]
\mynewtheorem{cor}{Corollary}%[section]
\mynewtheorem{prop}{Proposition}%[section]
\mynewtheorem{exa}{Example}%[section]
\mynewtheorem{alg}{Algorithm}%[section]
\mynewtheorem{rem}{Remark}%[section]
\mynewtheorem{bsp}{Example}

\def\equationautorefname~#1\null{Equation~(#1)\null}

\newcommand{\aref}[1]{\hyperref[#1]{Appendix~\ref{#1}}}

\usepackage{sectsty}
\allsectionsfont{\sffamily}

\usepackage{url}

\usepackage{caption}
\usepackage{footnote}

\newcommand{\Rbb}{\mathbb{R}}

\newcommand{\thetab}{\boldsymbol{\theta}}
\newcommand{\ub}{\mathbf{u}}
\newcommand{\Ub}{\mathbf{U}}

\newcommand{\vb}{\mathbf{v}}

\newcommand{\Xb}{\mathbf{X}}
\newcommand{\xb}{\mathbf{x}}

\newcommand{\yb}{\mathbf{y}}

\newcommand{\Nc}{\mathcal{N}}

\newcommand{\be}{\begin{equation}}
\newcommand{\ee}{\end{equation}}

\newcommand{\eps}{\varepsilon}

\DeclareMathOperator{\Var}{Var}

\DeclareMathOperator{\AIC}{AIC}
\DeclareMathOperator{\cll}{cll}

\DeclareMathOperator{\MRASE}{MRASE}
\DeclareMathOperator*{\argmin}{argmin} % replaced
\DeclareMathOperator{\ran}{ran}
\DeclareMathOperator{\SNR}{SNR}
\newcommand{\param}{\bm{\theta}}
\newcommand{\al}{{\alpha}}
\usepackage{bm}
\usepackage{mathtools}
\usepackage[overload]{empheq}
\usepackage{array}
\usepackage{caption}
\newcolumntype{C}[1]{>{\centering\arraybackslash}m{#1}}
\begin{document}
	
		\title{\bfseries \sffamily D-vine quantile regression with discrete variables}
			
%			\date{\small \today}
			\author{Niklas Schallhorn \and Daniel Kraus\footnote{Corresponding author: \href{mailto:daniel.kraus@tum.de}{daniel.kraus@tum.de}} \and Thomas Nagler \and Claudia Czado}
\date{%
	Zentrum Mathematik, Technische Universit\"at M\"unchen\\[2ex]%
	\today
}
			\maketitle
\vspace*{-0.2cm}
	\begin{abstract}
Quantile regression, the prediction of conditional quantiles, finds applications in various fields. Often, some or all of the variables are discrete. The authors propose two new quantile regression approaches to handle such mixed discrete-continuous data. Both of them generalize the continuous D-vine quantile regression, where the dependence between the response and the covariates is modeled by a parametric D-vine. D-vine quantile regression provides very flexible models, that enable accurate and fast predictions. Moreover, it automatically takes care of major issues of classical quantile regression, such as quantile crossing and interactions between the covariates. The first approach keeps the parametric estimation of the D-vines, 
but modifies the formulas to account for the discreteness. The second approach estimates the D-vine using continuous convolution to make the discrete variables continuous and then estimates the D-vine nonparametrically. A simulation study is presented examining for which scenarios the discrete-continuous D-vine quantile regression can provide superior prediction abilities. Lastly, the functionality of the two introduced methods is demonstrated by a real-world example predicting the number of bike rentals.
	\end{abstract}
	\noindent \textit{Keywords:} quantile regression; discrete variables; continuous convolution; nonparametric; vine copulas 	
%\vspace{1cm}
\section{Introduction}
Quantile regression, the estimation of quantiles of a response random variable conditioned covariates, has gained importance in various fields since its first appearance in \cite{Koenker78}.
\cite{kraus2017d} propose a new method of quantile regression, where the dependence between the response and the covariates is modeled by a parametric D-vine as introduced in \cite{Aas09}.
The D-vine is estimated by sequentially adding variables to the model until none of the remaining variables provides additional information.
The D-vine approach remedies various shortcoming of classical quantile regression.
The models are flexible and parsimonious, they prevent quantile crossing, and interactions between covariates are automatically taken into account.
\cite{kraus2017d} show that the D-vine quantile regression is a competitive approach that often shows superior prediction quality.

However, the model proposed by \cite{kraus2017d} requires that the marginal distributions of the response and all of the covariates are continuous.
\cite{Genest07} give an overview of the difficulties of copula models with discrete variables.
Implications are, for instance, that the copula is no longer uniquely defined and that the dependence between variables is not captured by the copula alone, but also involves the discrete marginal distributions.
Taking these implications into account, \cite{Panagiotelis12} present an algorithm to fit vine copula models to purely discrete data.
\cite{Onken16} modify this algorithm to allow for cases where only some of the variables are discrete.
Quantile regression methods that can handle discrete data are, e.g., linear quantile regression  \citep{Koenker78}, additive quantile regression \citep{Koenker11,Fenske12}, and kernel quantile regression \citep{Li13}.

In this paper, we modify the continuous D-vine quantile regression from \cite{kraus2017d} in two ways. The first extends the formulas of the parametric model of \cite{kraus2017d} such that it can handle mixed discrete-continuous data. In contrast to the purely continuous setting, the conditional quantiles cannot be expressed in closed form but can be calculated by numerically inverting the conditional distribution function.
The second approach replaces the parametric estimation of pair-copulas by a nonparametric kernel density estimator. Discrete variables are handled by adding a small amount of noise which makes them continuous. As shown by \citet{nagler2017generic}, the resulting estimator is still a valid estimator of the discrete-continuous conditional quantile function.
Thereby, the simplicity of the continuous D-vine quantile regression is preserved in the discrete-continuous setting when using a nonparametric estimation approach.

The remainder of the paper is organized as follows.
\autoref{cha2} introduces the continuous D-vine quantiles regression including the necessary concepts of D-vine copulas,
while \autoref{cha3} presents the two approaches described above to handle discrete data.
A simulation study that compares the two discussed methods to several competitor methods is shown in \autoref{cha4}.
\autoref{cha5} applies the proposed methods to a real-world example of bike rentals. Finally, \autoref{cha6} draws conclusions and gives an outlook to areas of further research.

\section{Parametric D-vine quantile regression for continuous variables}
\label{cha2}
This section is a summary of what is explained in more detail in Sections 2 and 3 of \cite{kraus2017d}. We are interested in the conditional quantiles $q_{\alpha}$ at some quantile level $\alpha$ of a continuous response $Y$ given a continuous covariate vector $\Xb=(X_1,\ldots,X_d)'$ taking on values $\xb=(x_1,\ldots,x_d)'$. The conditional quantile is defined as the inverse of the conditional distribution function of $Y$ given $\Xb$, i.e.\
\begin{equation}
q_{\alpha}(x_1,\ldots,x_d):=F^{-1}_{Y|X_1,\ldots,X_d}(\alpha|x_1,\ldots,x_d).
\label{eq:defi_q_alpha}
\end{equation}
Using Sklar's Theorem \citep{sklar1959fonctions} the right hand-side can be expressed in terms of the marginal distributions $F_Y$ of $Y$ and $F_j$ of $X_j$ and the copula between $Y$ and $\Xb$ as
\begin{equation}
F^{-1}_{Y|X_1,\ldots,X_d}(\alpha|x_1,\ldots,x_d)=F^{-1}_Y\left(C^{-1}_{V|U_{1},\ldots,
	U_{d}}(\alpha|u_{1},\ldots,u_{d})\right),
\label{eq:cond_quantile}
\end{equation}
where $V=F_Y(Y)$, $U_j=F_j(X_j)$ are the uniformly distributed probability integral transforms of $Y$ and $\Xb$, and $u_j=F_j(x_j)$ are their realizations. Further, $C_{V,U_1,\ldots,U_d}$ is the distribution function of $(V, U_1, \ldots, U_d)^\prime$ and called the copula associated with the joint distribution of $Y$ and $\Xb$ \citep[for an introduction to copulas, see,][]{joe1997multivariate, nelsen2007introduction} and  $C_{V|U_1,\ldots,U_d}$ is the associated conditional distribution function of $V$ given $(U_1, \ldots, U_d)^\prime$. This representation facilitates flexible modeling of $q_{\alpha}$ plugging in suitable estimators for the marginal distributions and the copula. \cite{kraus2017d} propose to use kernel estimators for the marginals and a simplified D-vine copula for modeling $C_{V,\Ub}$. A simplified D-vine copula is a special case of regular vine copulas \citep[see][]{Aas09,bedford2002vines}. It constructs a $d$-dimensional copula density using the product of conditional and unconditional bivariate copulas:
\begin{multline}
c(u_1,\ldots,u_d) = \prod_{i=1}^{d-1}\prod_{j=i+1}^dc_{ij;i+1,\ldots,j-1}\big(C_{i|i+1,\ldots,j-1}\left(u_i|u_{i+1},\ldots,u_{j-1}\right),\\
C_{j|i+1,\ldots,j-1}\left(u_j|u_{i+1},\ldots,u_{j-1}\right)\big).
\label{eq:D-vine_density}
\end{multline}
Note that due to the simplifying assumption the pair-copulas $c_{ij;i+1,\ldots,j-1}$ do not depend on the conditioning values $u_{i+1},\ldots,u_{j-1}$ \citep[see][for a more detailed discussion of the simplifying assumption]{haff2010simplified,stoeber2013simplified,killiches2016examination}. The arguments $C_{i|D}\left(u_i|\ub_D\right)$ of the pair-copulas $c_{ij;D}$ can be derived recursively \citep{joe1997multivariate} using that for $l\in D$ and $D_{-l}:=D\backslash \left\{l\right\}$ it holds that
\begin{equation}
C_{i|D}\left(u_i|\ub_D\right) = h_{i|l;D_{-l}}\left(C_{i|D_{-l}}\left(u_i|\xb_{D_{-l}}\right)|C_{l|D_{-l}}\left(u_{l}|\ub_{D_{-l}}\right)\right),
\label{eq:Joe97}
\end{equation}
where $h_{i|j;D}(u|v)={\partial C_{ij;D}(u,v)}/{\partial v}$ is called the \emph{h-function} associated with the pair-copula $C_{ij;D}$.\\
D-vine copulas inherit the great modeling flexibility attributed to vine copulas: every pair-copula can be modeled with a different copula family and parameter. The main reason why \cite{kraus2017d} use a D-vine copula model in \autoref{eq:cond_quantile} is that the inverse of the conditional distribution of the first variable $V$ given the covariates $\Ub$ can be expressed analytically as a recursion over h-functions and their inverses. For example, in three dimensions the recursion in \autoref{eq:Joe97} can be used to express the conditional distribution of $V$ given $(U_1,U_2)'$ as
\[
C_{V|U_1,U_2}(v|u_1,u_2)={h_{V|U_2;U_1}}({ h_{V|U_1}}(v|u_1)|{ h_{U_2|U_1}}(u_2|u_1)),
\]
and therefore the conditional quantile as
\[
C^{-1}_{V|U_1,U_2}(\alpha|u_1,u_2)= { h^{-1}_{V|U_1}}\left({ h^{-1}_{V|U_2;U_1}}\left(\alpha|{ h_{U_2|U_1}}( u_2| u_1)\right)\big| u_1\right).
\]
The order of the covariates in the D-vine can be chosen arbitrarily. \cite{kraus2017d} proposed a fitting algorithm that sequentially adds the covariate to the model that improve the model fit the most. The model fit is measured in terms of the conditional log-likelihood for $V$ given $\ub$. More precisely, given copula data $v_i$ and $\ub_i$ and a fitted D-vine copula density $\hat c_{V,\Ub}$ the conditional log-likelihood (cll) is defined by $\sum_{i=1}^{n}\log \hat c_{V|\Ub}(v_i|\ub_i)$. Here, $c_{V|\Ub}$ is the density associated with $C_{V|\Ub}$. Variables that do not improve the model fit are omitted, thus accomplishing an automatic forward covariate selection. Depending on the desired degree of parsimony, instead of the cll one can also use an AIC- or BIC-corrected version of the cll, penalizing the number of parameters in the model \citep[cf.][]{kraus2017d}.\\
In a simulation study as well as real data applications, \cite{kraus2017d} demonstrate the superiority of D-vine quantile regression over competitor methods in many settings. However, two important issues of D-vine quantile regression were left as open research problems: the need for nonparametric  pair-copulas to avoid misspecifications as described in \cite{Dette14}, as well as the inability to handle data containing discrete variables. In the following section two new approaches are presented that generalize D-vine quantile regression to account for discrete data: one is parametric (\autoref{cha3.1}) and the other is nonparametric (\autoref{cha3.2}).

\section{D-vine quantile regression for mixed discrete and continuous variables}
\label{cha3}
Let now some or all of the variables, the response variable $Y$ and the predictors $X_j$, be discrete.
For $ j \in \{1,\dots,d\} $, let $ y \in \ran(F^{-1}_Y), \, x_j \in \ran(F^{-1}_j) $ be observed values on the original scale and $ v \coloneqq F_Y(y), \, u_j \coloneqq F_j(x_j) $ the associated values on the original scale. Here, $\ran(G^{-1})$ is the set of all possible values of a random variable with distribution $G$.
As before, we express the conditional quantile of $Y$ given $\Xb=\xb$ in the following way:
$$ q_{\al}(x_1,\dots,x_d) = F^{-1}_Y(C^{-1}_{V|U_1,\dots,U_d}(\al|u_1,\dots,u_d)). $$
To compute $ C^{-1}_{V|U_1,\dots,U_d}(\al|u_1,\dots,u_d) $, we need to model the joint distribution of $V$ and $U_1,\dots,U_d $.
Similar to the continuous case, this joint distribution is modeled using a D-vine that has $V$ fixed as the first node, which enables us to compute the conditional quantiles in an easy fashion. 

\subsection{Parametric modeling}
\label{cha3.1}
Let $ u,v, u_1, u_2, v_1, v_2 \in [0,1] $, $ u_1 > u_2, v_1 > v_2 $.
As analogues of the h-functions $ h_{i|j;D}(u,v) = \partial C_{i,j;D}(u,v) / \partial v $ for continuous conditioning variables, we define for discrete conditioning variables with $ i,j \notin D $,
\begin{subequations}\label{eq:htilde}
	\begin{align}
		\tilde{h}_{i|j;D}(u|v_1,v_2) &\coloneqq \frac{C_{i,j;D}(u,v_1) - C_{i,j;D}(u,v_2)}{v_1-v_2}, \\
		\tilde{h}_{j|i;D}(v|u_1,u_2) &\coloneqq \frac{C_{i,j;D}(u_1,v) - C_{i,j;D}(u_2,v)}{u_1-u_2}.
	\end{align}
\end{subequations}
For a discrete random variable $X$, $ x \in \ran(F^{-1}_X) $ and $ u = F_X(x) $, we denote by $ u^- \coloneqq F_X(x^-) \coloneqq \lim_{a \nearrow x} F_X(a) $ the PIT-value of the next smaller value attained by $X$.

The following expressions for the conditional distribution function and the conditional density are derived in \cite{Stoeber13}.
More detailed derivations including explicit expressions for all cases in 2 and 3 dimensions can be found in Chapter 3 of \cite{Schallhorn17}.

If the joint distribution of $ (V,\Ub) $ is modeled by a D-vine with order $ V-U_{l_1}-\ldots-U_{l_d} $, with $ (l_1,\dots,l_d)' $ being a permutation of $ (1,\dots,d) $, 
then the conditional distribution function $ C_{V|\Ub}(v|\ub) $ can be computed iteratively by
\begin{subequations}
	\begingroup
	\begin{flalign}[left ={C_{V|\Ub}(v|\ub)= \empheqlbrace}]
	    	& h_{V|U_{l_d};\Ub_{-l_d}}\big(C_{V|\Ub_{-l_d}}(v|\ub_{-l_d}) \big| C_{U_{l_d}|\Ub_{-l_d}}(u_{l_d}|\ub_{-l_d})\big), \hspace{0.5cm} X_{l_d} \text{ continuous}, \\[0.8em]
    		& \tilde{h}_{V|U_{l_d};\Ub_{-l_d}}\big(C_{V|\Ub_{-l_d}}(v|\ub_{-l_d}) \big| C_{U_{l_d}|\Ub_{-l_d}}(u_{l_d}|\ub_{-l_d}),  \nonumber \\
		&		\hspace{4.82cm} C_{U_{l_d}|\Ub_{-l_d}}(u^-_{l_d}|\ub_{-l_d})\big), \hspace{0.5cm} X_{l_d} \text{ discrete}.
	\end{flalign}
	\endgroup
\end{subequations}
For example, when $X_1$ is discrete we have
\begin{align*}
	C_{V|U_1}(v|u_1) &= \frac{Pr(V \leq v, U_1 = u_1)}{Pr(U_1=u_1)} = \frac{Pr(V \leq v, U_1 \leq u_1) - Pr(V \leq v, U_1 \leq u_1^-)}{Pr(U_1 \leq u_1) - Pr(U_1 \leq u_1^-)} \\
	& = \frac{F_{V,U_1}(v,u_1) - F_{V,U_1}(v,u_1^-)}{u_1-u_1^-} = \frac{C_{V,U_1}(F_V(v),F_{U_1}(u_1)) - C_{V,U_1}(F_V(v),F_{U_1}(u_1^-))}{u_1-u_1^-} \\
	& = \frac{C_{V,U_1}(v,u_1) - C_{V,U_1}(v,u_1^-)}{u_1-u_1^-} = \tilde{h}_{V|U_1}(v|u_1,u_1^-).
\end{align*}

For the estimation of the D-vine, we need to estimate pair-copulas in a discrete-continuous setting.
If $C_{1,2}$ is the pair-copula of $U_1$ and $U_2$, then their joint density is given by
\begin{subequations}
	\begingroup
	\begin{flalign}[left ={f_{U_1,U_2}(u_1,u_2; C_{1,2}) = \empheqlbrace}]
	    	& c_{1,2}(u_1,u_2), \hspace{3.55cm} X_1, X_2 \text{ both continuous}, \\[0.8em]
    		& h_{2|1}(u_2|u_1) - h_{2|1}(u^-_2|u_1), \hspace{1.2cm} X_1 \text{ continuous}, X_2 \text{ discrete}, \\[0.8em]
		& h_{1|2}(u_1|u_2) - h_{1|2}(u^-_1|u_2), \hspace{1.2cm} X_1 \text{ discrete}, X_2 \text{ continuous}, \\[0.8em]
		& C_{1,2}(u_1,u_2) - C_{1,2}(u^-_1,u_2) \, - \nonumber \\
		& C_{1,2}(u_1,u^-_2) + C_{1,2}(u^-_1,u^-_2), \hspace{0.8cm} X_1,X_2 \text{ both discrete}.
	\end{flalign}
	\endgroup
	\label{eq1}
\end{subequations}
For given data $ \xb_j = \big(x_j^{(1)},\dots,x_j^{(n)}\big) $, estimated marginal distribution functions $ \hat{F}_j $ and $ \hat{\ub}_j \coloneqq \hat{F}_j(\xb_j) $ for $ j=1,2 $,
the copula $C_{1,2}$ is estimated parametrically by minimizing the AIC
\begin{equation} \AIC(C_{1,2}; \hat{\ub}_1, \hat{\ub}_2) \coloneqq -2 \sum_{i=1}^n \log f_{U_1,U_2}\Big(\hat{u}_1^{(i)},\hat{u}_2^{(i)}; C_{1,2}\Big) + 2 |\param(C_{1,2})| \end{equation}
over all available pair-copula families and their parameters, where $ |\param(C_{1,2})| $ is the number of parameters of the pair-copula $ C_{1,2} $.

For independent observations $ \yb=\big(y^{(1)},\dots,y^{(n)}\big), \, \xb_j=\big(x_j^{(1)},\dots,x_j^{(n)}\big) $, $ j=1,\dots,d $ of the variables $Y$ and $ X_1,\dots,X_d$,
we set $ \hat{\vb} \coloneqq \hat{F}_Y(\yb) $ and $ \hat{\ub}_j \coloneqq \hat{F}_j(\xb_j) $ using the estimated marginal distribution functions.
To fit the D-vine, the same estimation process of sequentially adding variables to the D-vine as for the continuous case is used.
However, the conditional density $ f_{V|\Ub} $ in the conditional log-likelihood (cll) is computed recursively as follows. 
\begingroup
\addtolength{\jot}{0.1em}
\begin{subequations}
	\text{If both $Y$ and $X_{l_d}$ are continuous, then it holds}
	\begin{flalign}
		f_{V|\Ub}&\Big(\hat{v}^{(i)} \Big| \hat{\ub}^{(i)}\Big) = & \nonumber \\
		& c_{V,U_{l_d};\Ub_{-l_d}}\Big( F_{V|\Ub_{-l_d}}\big(\hat{v}^{(i)}|\hat{\ub}_{-l_d}^{(i)}\big), F_{U_{l_d}|\Ub_{-l_d}}\big(\hat{u}_{l_d}^{(i)}|\hat{\ub}_{-l_d}^{(i)}\big) \Big) \cdot f_{V|\Ub_{-l_d}}\Big(\hat{v}^{(i)} \Big| \hat{\ub}_{-l_d}^{(i)}\Big).
	\end{flalign}
	\text{If $Y$ is continuous and $X_{l_d}$ is discrete, then the following is fulfilled:}
	\begin{flalign}
		f_{V|\Ub}\Big(\hat{v}^{(i)} \Big| \hat{\ub}^{(i)}\Big) = & \Big[ h_{U_{l_d}|V;\Ub_{-l_d}}\Big( F_{U_{l_d}|\Ub_{-l_d}}\big(\hat{u}_{l_d}^{(i)}|\hat{\ub}_{-l_d}^{(i)}\big) \Big| F_{V|\Ub_{-l_d}}\big(\hat{v}^{(i)}|\hat{\ub}_{-l_d}^{(i)}\big) \Big) & \nonumber \\
			& - h_{U_{l_d}|V;\Ub_{-l_d}}\Big( F_{U_{l_d}|\Ub_{-l_d}}\big(\hat{u}_{l_d-}^{(i)}|\hat{\ub}_{-l_d}^{(i)}\big) \Big| F_{V|\Ub_{-l_d}}\big(\hat{v}^{(i)}|\hat{\ub}_{-l_d}^{(i)}\big) \Big) \Big] \nonumber \\
			& \cdot \frac{f_{V|\Ub_{-l_d}}\Big(\hat{v}^{(i)} \Big| \hat{\ub}_{-l_d}^{(i)}\Big)}{F_{U_{l_d}|\Ub_{-l_d}}\big(\hat{u}_{l_d}^{(i)}|\hat{\ub}_{-l_d}^{(i)}\big) - F_{U_{l_d}|\Ub_{-l_d}}\big(\hat{u}_{l_d-}^{(i)}|\hat{\ub}_{-l_d}^{(i)}\big)}.
	\end{flalign}
	\text{If $Y$ is discrete and $X_{l_d}$ is continuous, we have}
	\begin{flalign}
		f_{V|\Ub}\Big(\hat{v}^{(i)} \Big| \hat{\ub}^{(i)}\Big) = \; & h_{V|U_{l_d};\Ub_{-l_d}}\Big( F_{V|\Ub_{-l_d}}\big(\hat{v}^{(i)}|\hat{\ub}_{-l_d}^{(i)}\big) \Big| F_{U_{l_d}|\Ub_{-l_d}}\big(\hat{u}_{l_d}^{(i)}|\hat{\ub}_{-l_d}^{(i)}\big) \Big) & \nonumber \\
			& - h_{V|U_{l_d};\Ub_{-l_d}}\Big( F_{V|\Ub_{-l_d}}\big(\hat{v}_-^{(i)}|\hat{\ub}_{-l_d}^{(i)}\big) \Big| F_{U_{l_d}|\Ub_{-l_d}}\big(\hat{u}_{l_d}^{(i)}|\hat{\ub}_{-l_d}^{(i)}\big) \Big).
	\end{flalign}
	\text{If both $Y$ and $X_{l_d}$ are discrete, it holds}
	\begin{flalign}
		f_{V|\Ub}&\Big(\hat{v}^{(i)} \Big| \hat{\ub}^{(i)}\Big) = & \nonumber \\
		& \tilde{h}_{V|U_{l_d};\Ub_{-l_d}}\Big( F_{V|\Ub_{-l_d}}\big(\hat{v}^{(i)}|\hat{\ub}_{-l_d}^{(i)}\big) \Big| F_{U_{l_d}|\Ub_{-l_d}}\big(\hat{u}_{l_d}^{(i)}|\hat{\ub}_{-l_d}^{(i)}\big),	F_{U_{l_d}|\Ub_{-l_d}}\big(\hat{u}_{l_d-}^{(i)}|\hat{\ub}_{-l_d}^{(i)}\big) \Big) \nonumber \\
		& - \tilde{h}_{V|U_{l_d};\Ub_{-l_d}}\Big( F_{V|\Ub_{-l_d}}\big(\hat{v}_-^{(i)}|\hat{\ub}_{-l_d}^{(i)}\big) \Big| F_{U_{l_d}|\Ub_{-l_d}}\big(\hat{u}_{l_d}^{(i)}|\hat{\ub}_{-l_d}^{(i)}\big), F_{U_{l_d}|\Ub_{-l_d}}\big(\hat{u}_{l_d-}^{(i)}|\hat{\ub}_{-l_d}^{(i)}\big) \Big).
	\end{flalign}
\end{subequations}
\endgroup
Here, we set $ \hat{v}_-^{(i)} \coloneqq \big(\hat{v}^{(i)}\big)^- $ and $ \hat{u}_{l_d-}^{(i)} \coloneqq \big(\hat{u}_{l_d}^{(i)}\big)^- $.

Once the D-vine is specified, then $ C^{-1}_{V|U_1,\dots,U_d}(\al|u_1,\dots,u_d) $ can be obtained by numerically inverting $ C_{V|\Ub}(\cdot|\ub) \coloneqq C_{V|U_1,\dots,U_d}(\cdot|u_1,\dots,u_d) $, i.e.\
\begin{equation} C^{-1}_{V|\Ub}(\al|\ub) = \argmin_{\mathclap{\substack{q \in [0,1] \\ C_{V|\Ub}(q|\ub) \geq \al}}}~ \Big(C_{V|\Ub}(q|\ub) - \al \Big), \end{equation}
opposed to the continuous case, where $ C^{-1}_{V|\Ub}(\al|\ub) $ is expressed in terms of nested inverse h-functions.
This modification is made since the $ \widetilde{\text{h}} $-functions defined in \autoref{eq:htilde} would need to be inverted numerically.
Hence, it is more stable to directly numerically invert the conditional distribution function composed of nested h- and $ \widetilde{\text{h}} $-functions instead of computing the conditional quantile function using several numerically inverted $ \widetilde{\text{h}} $-functions.

To ensure that non-influential variables are excluded and to reflect preference for parsimonious models, we use the AIC-corrected conditional log-likelihood as defined for the continuous case \citep{kraus2017d}, i.e.\ $\cll_{\AIC}=-2\cll+2|\thetab|$, where $|\thetab|$ is the number of parameters of the fitted D-vine. Since the pair-copulas in the estimation of the D-vine are determined parametrically, we call this method \textbf{parametric D-vine quantile regression (PDVQR)}.

\subsection{Nonparametric modeling}
\label{cha3.2}
Vine copula models can also be estimated nonparametrically. For the case where all variables are continuous, \citet{nagler2017nonparametric} surveyed existing methods for estimation of the vine copula density. It is straightforward to use these methods as nonparametric D-vine quantile regression estimators by following the construction of \citet{kraus2017d}: Given an estimate $\widehat c_{V, U_1, \dots, U_d}$ of the joint density of $(V, U_1, \dots, U_d)$, we can derive an estimate of the conditional distribution function $C_{V | U_1, \dots, U_d}$ as
\begin{align*}
\widehat C_{V| U_1, \dots, U_d}(v\vert u_1, \dots, u_d) = \int_0^v \widehat c_{V| U_1, \dots, U_d}(s\vert u_1, \dots, u_d) ds.
\end{align*}
A nonparametric estimator of the conditional quantile function $q_{a}$ is then defined by invoking \eqref{eq:cond_quantile}. In the continuous case, $C^{-1}_{V | U_1, \dots, U_d}$ can even be derived in (almost) closed form, only involving h-functions and their inverses  \citep[see,][]{kraus2017d}.

This construction is straightforward as long as all variables are continuous, but none of the methods in \citet{nagler2017nonparametric} are applicable when some of the variables are discrete. Furthermore, the arguments in \autoref{cha3.1} do not apply since they are specific to maximum likelihood inference of a finite-dimensional parameter. Our solution to this problem is based on \emph{continuous convolution}. The idea is to make all discrete variables continuous by adding a small amount of noise. \citet{nagler2017generic} showed that this still leads to valid estimators of conditional quantile functions if the noise distribution belongs to a certain class. We shall make this more precise in the following paragraphs.

For $\mathcal{D} \subseteq \{1, \dots, d\}$, suppose that $Y$ and $X_j$, $j \in \{1, \dots, d\} \setminus \mathcal{D},$ are continuous variables, whereas $X_j, j \in \mathcal{D},$ are discrete. Let further $E_j, j \in \mathcal{D}$, be \emph{iid} random variables independent of $(Y, X_1, \dots, X_d)$ with density $\eta$ satisfying the following constraint: for some $0 < \gamma_1 \le \gamma_2 < 1$, $\eta(x) = 1$ for $x \in [-\gamma_1, \gamma_1]$ and $\eta(x) = 0$ for $x \in \Rbb \setminus (-\gamma_2, \gamma_2)$. An example of such a density is $\eta(x) = \mathds{1}(\vert x \vert < 0.5)$, i.e., the $E_j$'s are uniformly distributed on $(-0.5, 0.5)$.

The continuous convolution of $(X_1, \dots, X_d)$ is defined as $(\tilde X_1, \dots, \tilde X_d)$, where $\tilde X_j = X_j + E_j$ for all $j \in \mathcal{D}$, and $\tilde X_j = X_j$ for all $j \in \{1, \dots, d\} \setminus \mathcal{D}$. Then Proposition 5 of \citet{nagler2017generic} shows that for all $\alpha \in [0, 1]$, $(x_1, \dots, x_d) \in \bigtimes_{j=1}^d \mathrm{ran}(X_j)$,
\begin{align} \label{eq:cc_equality}
F^{-1}_{Y|X_1,\ldots,X_d}(\alpha|x_1,\ldots,x_d) = F^{-1}_{Y|\tilde X_1,\ldots,\tilde X_d}(\alpha|x_1,\ldots,x_d). 
\end{align}
The right hand side of \eqref{eq:cc_equality} is the conditional quantile function of continuous variables only. It can thus be estimated by using any of the nonparametric methods in \citet{nagler2017nonparametric}. And since \eqref{eq:cc_equality} is an equality, this also yields an estimator of the left hand side, the conditional quantile function we are actually interested in. The case where $Y$ is discrete can be handled similarly. However, a correction term has to be added to the right hand side of \eqref{eq:cc_equality} \citep[see,][Proposition 5]{nagler2017generic}. \citet{nagler2017generic} stressed that this approach is only valid for nonparametric estimation and, thus, must not be used with parametric models.

In the context of density estimation, \citet{Nagler2016} showed that nonparametric estimators based on simplified vine copulas  have an appealing property: they do not suffer from the curse of dimensionality. More specifically, convergence rates are equivalent to those of a two-dimensional problem, no matter how large $d$ actually is. Since the D-vine quantile regression estimator is derived from the estimated density, similar findings can be expected in our setting. The exact asymptotic behavior can be established by arguments similar to those in \citet{Nagler2016}, but is beyond the scope of this article. 

In the simulations and application we will use the vine copula density estimator that performed best in \citet{nagler2017nonparametric}. It estimates the pair-copula densities by a local likelihood approach proposed by \citet{geenens2017}. For the noise density $\eta$, we choose the uniform density, $\eta(x) = \mathds{1}(\vert x \vert < 0.5)$. We call this method {\bfseries nonparametric D-vine quantile regression (NPDVQR)}.

\section{Simulation study}
\label{cha4}
We will compare the two methods presented in this paper to three other commonly used methods for quantile regression.
We start by a brief summary of the competitor methods, followed by a description of the simulation setup and results.

\subsection{Competitor methods}
\textbf{Linear quantile regression (LQR)} \quad
Introduced in \cite{Koenker78}, it is assumed that the conditional quantiles linearly depend on the conditioning values, i.e.
$$ \hat{q}_\al(x_1,\dots,x_d) = \hat{\beta}_0 + \sum_{j=1}^d \hat{\beta}_j x_j. $$
The estimates for the regression coefficients $ \hat{\beta}_j $ are obtained as the solution of the minimization problem
$$ \min_{\beta \in \Rbb^d} \Bigg( \al \sum_{i=1}^n \bigg(y^{(i)} - \beta_0 - \sum_{j=1}^d \beta_j x_j^{(i)}\bigg)^+ + (1-\al) \sum_{i=1}^n \bigg(\beta_0 + \sum_{j=1}^d \beta_j x_j^{(i)} - y^{(i)}\bigg)^+ \Bigg). $$
This method has various shortcomings described in \cite{Bernard15}, for instance the estimates are not necessarily monotonically increasing in $\al$.
LQR can be performed using the function \texttt{rq} of the package \texttt{quantreg} \citep{Koenker15}. \vspace{0.3cm}

\textbf{Boosted additive quantile regression (BAQR)} \quad
To relax the linear assumption as above, \cite{Koenker05} proposes to use additive models for quantile regression, i.e.
$$ \hat{q}_\al(x_1,\dots,x_d) = \hat{\beta}_0 + \sum_{j=1}^J \Bigg( \sum_{k=1}^{K_j-1} \hat{\beta}^k_j I^k_j(x_j) \Bigg) + \sum_{j=J+1}^d g_j(x_j), $$
where the discrete variables $ X_1,\dots,X_J $ are estimated by ordinary least squares using a dummy coding with $K_j$ denoting the number of values attained by $X_j$, and $g_j$ denotes a smooth function based on B-splines.
\cite{Fenske12} use a boosting technique to estimate the model parameters, minimizing a given loss function including penalizing terms by stepwise updating the estimator along the steepest gradient of the loss function.
The algorithm is implemented in the function \texttt{gamboost} of the package \texttt{mboost} \citep{Hothorn16}. \vspace{0.3cm}

\textbf{nonparametric quantile regression (NPQR)} \quad
As introduced in \cite{Li13}, the conditional quantiles are obtained via numerical inversion of the conditional distribution function, i.e.
$$ \hat{q}_\al(x_1,\dots,x_d) = \argmin_{q \in \Rbb} \Big| \widehat{F}_{Y|X_1,\dots,X_d}(q|x_1,\dots,x_d) - \al \Big|. $$
The estimate $ \widehat{F}_{Y|X_1,\dots,X_{d-1}} $ is obtained nonparametrically using a kernel estimator with an automatic data-driven bandwidth selector.
NPQR can be performed using the function \texttt{npqreg} of the package \texttt{np} \citep{Hayfield08}.\\
The three methods can handle continuous and discrete predictors. 
However, if $Y$ is discrete then the estimated quantiles are not necessarily values actually attained by $Y$, so the obtained conditional quantiles have to be rounded to the closest value attained by $Y$.

\subsection{Setup}
We compare the five methods in different settings.
For each setting and each replication $ r=1,\dots,R=100 $, we simulate a training dataset $ (y_{r,i}^{train},\xb_{r,i}^{train})_{i=1,\dots,n_{train}} $ from the joint distribution of $ (Y,\Xb) $
and an evaluation dataset $ (\xb_{r,i}^{eval})_{i=1,\dots,n_{eval}} $, $ n_{eval} = 1000 $, from the distribution of $\Xb$.
For each method $m$ and $ \al \in (0,1) $, we compute the estimate of the conditional quantile function $ \hat{q}_{m,\al}(\cdot) $ based on the training dataset.
The evaluation dataset is used to estimate the root average squared error.
We take the mean over all replications, giving us the out-of-sample mean root average squared error ($ {\MRASE}_{m,\al} $) of method $m$,
\begin{equation} {\MRASE}_{m,\al} \coloneqq \frac1R \sum_{r=1}^{R} \sqrt{ \frac{1}{n_{eval}} \sum_{i=1}^{n_{eval}} \Big( \hat{q}_{m,\al}\big(\xb_{r,i}^{eval}\big) - q_{\al}\big(\xb_{r,i}^{eval}\big) \Big)^2 }, \end{equation}
where $ q_{\al}(\cdot) $ denotes the true conditional quantile function.

The data is generated as follows.
Using the package \texttt{copula} \citep{Hofert16}, we simulate $ (\ub_1,\dots,\ub_d) $ from a d-dimensional Clayton copula with parameter $ \theta=1 $, corresponding to an unconditional pairwise Kendall's $\tau$ of $1/3$, and sample size $ n_{train} $, i.e.\ we have $ \ub_j \in [0,1]^{n_{train}} $.
We consider $ d \in \{3,5\} $ and $ n_{train} \in \{250,1000\} $.
The first two variables are discretized by applying the quantile function of the binomial distribution $ F^{-1}(\cdot;N,1/2) $ with parameters $ N \in \{2,8\} $ and $ p=1/2 $.
So if the $j$-th variable shall be discretized, we set
	$$ \xb_j = F^{-1}(\ub_j;N,1/2). $$
The remaining continuous variables are transformed using the quantile function of the standard normal distribution $\Phi^{-1}$,
i.e.\ if the $j$-th variable shall be continuous, we set
	$$ \xb_j = \Phi^{-1}(\ub_j). $$
We then compute 
$$ \yb = g(\xb_1,\dots,\xb_d) + \bm{\eps}, $$
with some function $ g:\Rbb^d \rightarrow \Rbb $ and $ \bm{\eps} \in \Rbb^{n_{train}} $ 
that consists of $ \eps_i = \sqrt{\frac{\Var(g(\Xb);\theta,N)}{\SNR}} \, Z_i $ with i.i.d.\ $ Z_i \sim \Nc(0,1) $.
Here, SNR denotes the signal-to-noise ratio, for which we consider $ \SNR \in \{0.5,2\} $, and $ \Var(g(\Xb);\theta,N) $ is the variance of $ g(\Xb) $ depending on $\theta$ and $N$.

\subsection{Results}

\autoref{tab1}, \autoref{tab2} and \autoref{tab3} present the results for the considered model specifications.
Additional results for other specifications can be found in \cite{Schallhorn17}.
For each model specification and each $\al$, the $ {\MRASE} $s marked in \textbf{bold} are the smallest $ {\MRASE} $ or those which are not significantly larger than the smallest $ {\MRASE} $.
The significance is measured by a t-test, for which we choose a significance level of 5\%.

\begin{table}[htbp]
	\centering
	\small
	\renewcommand{\arraystretch}{1.1}
	\captionsetup{font=normalsize}
	\begin{tabular}{ | C{0.9cm} | C{0.9cm} | C{0.9cm} | C{0.9cm} | C{1.6cm} | C{1.6cm} | C{1.6cm} | C{1.6cm} | C{1.6cm} | }
		\hline
		$\SNR$ & $n_{train}$ & $N$ & $\al$ & PDVQR & NPDVQR & LQR & BAQR & NPQR \\ \hline
		\multirow{8}{*}{0.5} & \multirow{4}{*}{250} & \multirow{2}{*}{2} & 0.01 & \textbf{1.27} & 1.88 & 1.62 & 1.82 & 1.73 \\ \cline{4-9}
		& & & 0.5 & 0.71 & 1.36 & \textbf{0.53} & 0.71 & 1.05 \\ \cline{3-9}
		& & \multirow{2}{*}{8} & 0.01 & \textbf{1.44} & 1.98 & 1.84 & 2.18 & 2.12 \\ \cline{4-9}
		& & & 0.5 & 0.82 & 1.02 & \textbf{0.58} & 0.78 & 1.40 \\ \cline{2-9}
		& \multirow{4}{*}{1000} & \multirow{2}{*}{2} & 0.01 & 1.08 & 1.01 & \textbf{0.87} & 1.48 & 1.29 \\ \cline{4-9}
		& & & 0.5 & 0.41 & 0.49 & \textbf{0.28} & 0.34 & 0.62 \\ \cline{3-9}
		& & \multirow{2}{*}{8} & 0.01 & 1.12 & 1.20 & \textbf{0.90} & 1.86 & 1.62 \\ \cline{4-9}
		& & & 0.5 & 0.56 & 0.56 & \textbf{0.31} & 0.46 & 0.90 \\ \cline{1-9}
		\multirow{8}{*}{2} & \multirow{4}{*}{250} & \multirow{2}{*}{2} & 0.01 & \textbf{0.82} & \textbf{0.84} & \textbf{0.81} & 1.01 & 1.05 \\ \cline{4-9}
		& & & 0.5 & 0.43 & 0.48 & \textbf{0.27} & 0.37 & 0.65 \\ \cline{3-9}
		& & \multirow{2}{*}{8} & 0.01 & \textbf{0.95} & 1.03 & \textbf{0.92} & 1.47 & 1.40 \\ \cline{4-9}
		& & & 0.5 & 0.55 & 0.58 & \textbf{0.29} & 0.41 & 0.93 \\ \cline{2-9}
		& \multirow{4}{*}{1000} & \multirow{2}{*}{2} & 0.01 & 0.79 & 0.62 & \textbf{0.44} & 0.81 & 0.77 \\ \cline{4-9}
		& & & 0.5 & 0.29 & 0.36 & \textbf{0.14} & 0.17 & 0.38 \\ \cline{3-9}
		& & \multirow{2}{*}{8} & 0.01 & 0.76 & 0.69 & \textbf{0.45} & 1.36 & 0.99 \\ \cline{4-9}
		& & & 0.5 & 0.40 & 0.35 & \textbf{0.15} & 0.20 & 0.57 \\
		\hline
	\end{tabular}
	\caption{${\MRASE}$ for $ d=3 $, linear $ g(\bm{x}_1,\bm{x}_2,\bm{x}_3) = 2 \bm{x}_1 - 3 \bm{x}_3 $ for all considered methods.}
	\label{tab1}
\end{table}

For \textbf{dimension 3 and $g$ linear} as shown in \autoref{tab1}, LQR has, as expected, the best prediction quality in almost every setting, 
while PDVQR clearly performs better for $ \al=0.01 $ in the cases with both large errors ($ \SNR = 0.5 $) and a small sample size ($ n_{train} = 250 $).
For a larger sample size and smaller errors however, LQR also performs better in the tails.
BAQR performs relatively bad in the tails, particularly when the errors are large, but it provides reasonable predictions for the conditional median. The performances of NPDVQR and NPQR are relatively bad, as to be expected in this linear case. However, NPDVQR and NPQR outperform BAQR for $\alpha = 0.01$ and NPDVQR is better than PDVQR in the setting $N = 8$ and $n = 1000$.

\begin{table}[!b]
	\centering
	\small
	\renewcommand{\arraystretch}{1.1}
	\captionsetup{font=normalsize}
	\begin{tabular}{ | C{0.9cm} | C{0.9cm} | C{0.9cm} | C{0.9cm} | C{1.6cm} | C{1.6cm} | C{1.6cm} | C{1.6cm} | C{1.6cm} | }
		\hline
		$\SNR$ & $n_{train}$ & $N$ & $\al$ & PDVQR & NPDVQR & LQR & BAQR & NPQR \\ \hline
		\multirow{8}{*}{0.5} & \multirow{4}{*}{250} & \multirow{2}{*}{2} & 0.01 & \textbf{3.22} & \textbf{3.12} & 4.01 & 6.25 & \textbf{3.26} \\ \cline{4-9}
		& & & 0.5 & 1.98 & \textbf{1.81} & 2.04 & \textbf{1.85} & \textbf{1.85} \\ \cline{3-9}
		& & \multirow{2}{*}{8} & 0.01 & 6.52 & \textbf{3.96} & 8.04 & 9.63 & 5.08 \\ \cline{4-9}
		& & & 0.5 & 3.97 & \textbf{2.60} & 5.24 & 5.62 & 3.00 \\ \cline{2-9}
		& \multirow{4}{*}{1000} & \multirow{2}{*}{2} & 0.01 & \textbf{2.20} & 2.43 & 2.66 & 5.92 & 2.43 \\ \cline{4-9}
		& & & 0.5 & 1.57 & 1.61 & 1.69 & 1.46 & \textbf{1.23} \\ \cline{3-9}
		& & \multirow{2}{*}{8} & 0.01 & 5.10 & \textbf{2.91} & 6.93 & 9.51 & 3.72 \\ \cline{4-9}
		& & & 0.5 & 3.70 & 2.14 & 5.05 & 5.46 & \textbf{2.03} \\ \cline{1-9}
		\multirow{8}{*}{2} & \multirow{4}{*}{250} & \multirow{2}{*}{2} & 0.01 & 2.20 & 2.68 & 2.62 & 2.19 & \textbf{1.99} \\ \cline{4-9}
		& & & 0.5 & 1.55 & 1.65 & 1.71 & 1.38 & \textbf{1.25} \\ \cline{3-9}
		& & \multirow{2}{*}{8} & 0.01 & 6.58 & 3.62 & 9.67 & 5.41 & \textbf{3.28} \\ \cline{4-9}
		& & & 0.5 & 3.77 & 2.31 & 5.08 & 5.36 & \textbf{2.15} \\ \cline{2-9}
		& \multirow{4}{*}{1000} & \multirow{2}{*}{2} & 0.01 & \textbf{1.55} & 2.37 & 2.02 & 1.92 & \textbf{1.55} \\ \cline{4-9}
		& & & 0.5 & 1.28 & 1.46 & 1.59 & 1.19 & \textbf{0.84} \\ \cline{3-9}
		& & \multirow{2}{*}{8} & 0.01 & 6.14 & 2.79 & 9.48 & 5.34 & \textbf{2.41} \\ \cline{4-9}
		& & & 0.5 & 3.55 & 1.99 & 5.03 & 5.28 & \textbf{1.43} \\
		\hline
	\end{tabular}
	\caption{${\MRASE}$ for $ d=3 $, non-linear $ g(\bm{x}_1,\bm{x}_2,\bm{x}_3) = \bm{x}_1 - 2 (\bm{x}_2-3)^2 + 4 \sqrt{|\bm{x}_3|} $ for all considered methods.}
	\label{tab2}
\end{table}

For \textbf{dimension 3 and $g$ non-linear} as shown in \autoref{tab2}, NPQR is the best method when the errors are small ($ \SNR = 2 $).
In the case of large errors however, it provides bad results for $ \al = 0.01 $, where the D-vine methods show the best prediction ability.
Between PDVQR and NPDVQR, PDVQR performs better for the highly discrete cases with $ N=2 $, while NPDVQR performs better for the more continuous cases with $ N=8 $.
The non-linear $g$-function implies a non-monotonic relationship between the response variable and the predictors $X_2$ and $X_3$.
As \cite{Dette14} show, none of the popular parametric pair-copula families  can model a non-monotonic dependency, disadvantaging PDVQR.
Since the non-monotonicity is stronger for $ N = 8 $, PDVQR shows worse predictions in the tails in these cases.
LQR performs worse than PDVQR, NPDVQR and NPQR for almost all model specifications. Again, BAQR performs considerably worse than the other methods (especially in the tails and for large errors) suggesting that it might be prone to over-fitting.

For \textbf{dimension 5 and $g$ non-linear} as shown in \autoref{tab3}, NPDVQR shows superior predictions for almost all specifications. The second best quantile prediction method for this non-linear example is NPQR. BAQR performs well for the conditional median in some cases, while it shows very large estimation errors for $ \al = 0.01 $ and $ \SNR = 0.5 $.
This might be due to the interaction term in the $g$-function, since we do not include interaction terms in the BAQR model.
There are also non-monotonic dependencies in the data, explaining why PDVQR does not provide very accurate predictions.
PDVQR still shows better predictions than BAQR for $ \al = 0.01 $ and a similar prediction quality for  $ \al = 0.5 $.
Again, LQR performs worse than PDVQR, NPDVQR and NPQR for all model specifications.

\begin{table}[htbp]
\centering
\small
\renewcommand{\arraystretch}{1.1}
\captionsetup{font=normalsize}
\begin{tabular}{ | C{0.9cm} | C{0.9cm} | C{0.9cm} | C{0.9cm} | C{1.6cm} | C{1.6cm} | C{1.6cm} | C{1.6cm} | C{1.6cm} | }
	\hline
	$\SNR$ & $n_{train}$ & $N$ & $\al$ & PDVQR & NPDVQR & LQR & BAQR & NPQR \\ \hline
	\multirow{8}{*}{0.5} & \multirow{4}{*}{250} & \multirow{2}{*}{2} & 0.01 & 6.05 & \textbf{4.86} & 8.04 & 11.11 & 5.52 \\ \cline{4-9}
	& & & 0.5 & 3.71 & \textbf{3.18} & 5.29 & 3.48 & 4.16 \\ \cline{3-9}
	& & \multirow{2}{*}{8} & 0.01 & \textbf{6.21} & \textbf{6.22} & 7.86 & 11.00 & \textbf{6.11} \\ \cline{4-9}
	& & & 0.5 & \textbf{4.19} & 4.49 & 5.44 & \textbf{4.19} & 4.66 \\ \cline{2-9}
	& \multirow{4}{*}{1000} & \multirow{2}{*}{2} & 0.01 & 5.38 & \textbf{3.74} & 6.18 & 10.51 & 4.43 \\ \cline{4-9}
	& & & 0.5 & 2.84 & \textbf{2.58} & 4.88 & 2.99 & \textbf{2.65} \\ \cline{3-9}
	& & \multirow{2}{*}{8} & 0.01 & 6.02 & \textbf{4.54} & 6.28 & 10.48 & 4.87 \\ \cline{4-9}
	& & & 0.5 & 3.46 & \textbf{3.19} & 5.05 & 3.85 & \textbf{3.11} \\ \cline{1-9}
	\multirow{8}{*}{2} & \multirow{4}{*}{250} & \multirow{2}{*}{2} & 0.01 & 5.21 & \textbf{3.71} & 6.14 & 5.95 & 4.32 \\ \cline{4-9}
	& & & 0.5 & 3.12 & \textbf{2.70} & 4.94 & \textbf{2.69} & 3.15 \\ \cline{3-9}
	& & \multirow{2}{*}{8} & 0.01 & 5.60 & \textbf{4.60} & 6.39 & 6.49 & 4.91 \\ \cline{4-9}
	& & & 0.5 & 3.75 & \textbf{3.33} & 5.12 & \textbf{3.27} & 3.61 \\ \cline{2-9}
	& \multirow{4}{*}{1000} & \multirow{2}{*}{2} & 0.01 & 4.72 & \textbf{3.15} & 5.62 & 5.57 & 3.49 \\ \cline{4-9}
	& & & 0.5 & 2.36 & 2.37 & 4.85 & 2.52 & \textbf{2.12} \\ \cline{3-9}
	& & \multirow{2}{*}{8} & 0.01 & 5.35 & \textbf{3.61} & 5.92 & 6.26 & 3.95 \\ \cline{4-9}
	& & & 0.5 & 3.14 & \textbf{2.65} & 5.01 & 3.15 & \textbf{2.57}\\
	\hline
\end{tabular}
\caption{${\MRASE}$ for $ d=5 $, non-linear $ g(\bm{x}_1,\bm{x}_2,\bm{x}_3,\bm{x}_4,\bm{x}_5) = 3 \sqrt{\bm{x}_1} - \bm{x}_3^2 + (\bm{x}_4+1)^3 - \bm{x}_2 \cdot \bm{x}_5 $ for all considered methods.}
\label{tab3}
\end{table}

In conclusion, the D-vine quantile regression methods provide much better predictions than LQR in the cases with a non-linear $g$.
Further, NPQR performed best among all methods in the non-linear scenario for $d=3$, but NPDVQR shows better results when $d$ is increased. This may be due to the fact that the convergence rate of NPDVQR is constant in $d$ \citep[cf.][]{Nagler2016}, while NPQR suffers from the curse of dimensionality. BAQR appears to work better when the signal-to-noise ratio is high. For the scenarios with lower signal-to-noise ratio errors, BAQR shows very large estimation errors (particularly in the tails) and is inferior to the D-vine quantile regressions methods.
For a linear $g$, LQR outperforms all other methods as the assumption of linearity is fulfilled. Interestingly, the D-vine quantile regression delivers better results in the tails in settings with a small sample size and large errors.
Thus, the D-vine quantile regression shows its merits particularly in the difficult cases, i.e., when the signal is hard to detect and extreme quantiles are the target.

We also want to briefly discuss the \textbf{run-time} for the model fitting and prediction of the conditional quantiles. Computation times over all 100 repetitions with $ \al \in \{0.01,0.5\} $, are shown in \autoref{tab4}.  PDVQR clearly has the longest run-time. 
BAQR shows the second and third longest run-time for $ n_{train}=250 $ and $ n_{train} = 1{,}000 $, 
while NPQR shows the third and second longest run-time for $ n_{train}=250 $ and $ n_{train} = 1{,}000 $.  NPDVQR is the fastest among the more sophisticated methods. Due to it's simplicity LQR can be computed almost instantly and is several orders of magnitude much faster than the other methods.

PDVQR is much slower than NPDVQR since the estimation of the pair-copulas takes more time.
This is because parameters have to be estimated for several pair-copula families before the best fitting model can be selected; NPDVQR only estimates one nonparametric model. Another factor is that the likelihood of each pair-copula in PDVQR is more complex than in the continuous case (see \autoref{eq1}), involving differences of the copula distribution function (which is demanding for some families). 

\begin{table}[htbp]
\centering
\small
\renewcommand{\arraystretch}{1.1}
\captionsetup{font=normalsize}
\begin{tabular}{ | C{0.9cm} | C{1.6cm} | C{1.6cm} | C{1.6cm} | C{1.6cm} | C{1.6cm} | }
	\hline
	$n_{train}$ & PDVQR & NPDVQR & LQR & BAQR & NPQR \\ \hline
	250 & 162.79 & 11.86 & 0.02 & 38.17 & 35.25 \\
	1,000 & 466.53 & 22.75 & 0.03 & 80.57 & 409.81 \\
	\hline
\end{tabular}
\caption{Run-times in seconds of the different methods for $d=3$, non-linear $g$, $\SNR=0.5$ and $N=2$ (recorded on an 8-way Opteron with 16 cores, each with 2.0 GHz and 16 GB of memory).
}
\label{tab4}
\end{table}

%\vspace*{-0.5cm}
\section{Application}
\label{cha5}

\begin{figure}[htbp]
    	\centering
    	\includegraphics[trim=0cm 0cm 0cm 0cm,clip,width=0.49\textwidth]{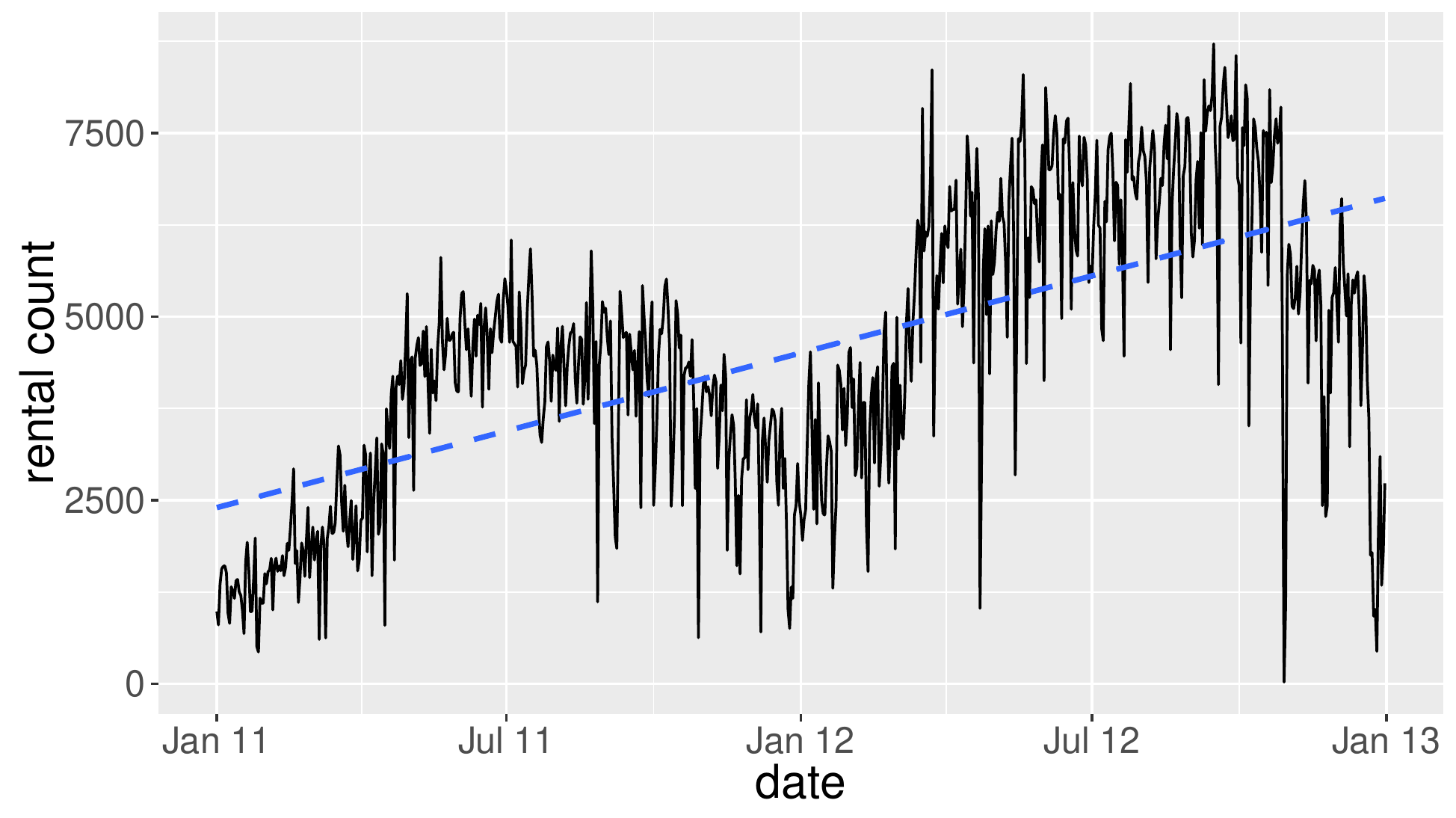}       
    	\includegraphics[trim=0cm 0cm 0cm 0cm,clip,width=0.49\textwidth]{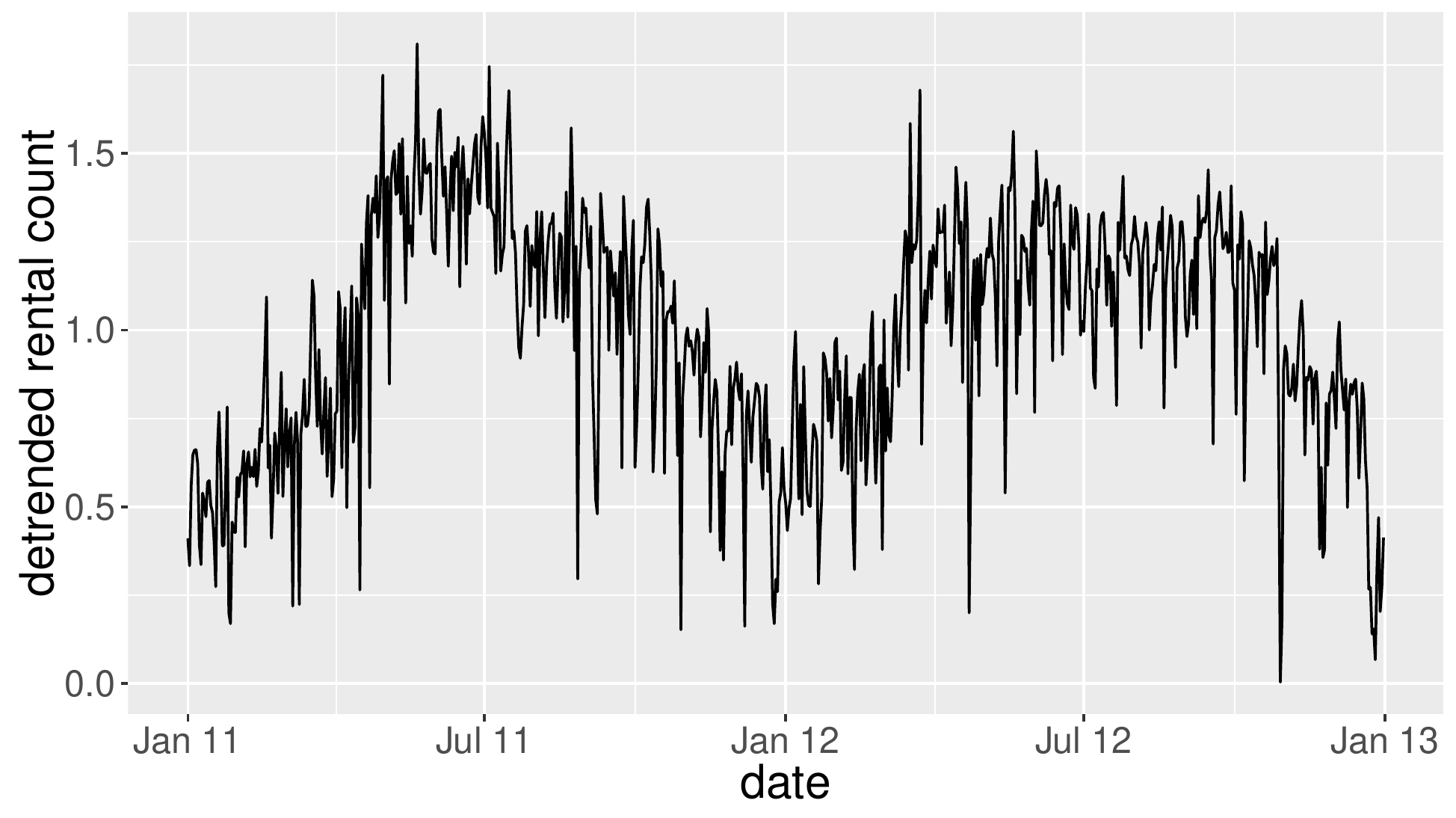}
    	\caption{Observed (left) and detrended (right) bike rental counts in the years 2011-2012. }
    	\label{fig:counts}
    \end{figure}

Thanks to the methods described in this paper, the application of D-vine quantile regression is no longer restricted to continuous data sets. We investigate the bike sharing data set from the UCI machine learning repository \citep{Lichman:2013}, first analyzed in \cite{fanaee2013event}. It contains information on rental counts from the bicycle sharing system \textit{Capital Bikeshare} offered in Washington, D.C., together with weather and seasonal information. As a response for the quantile regression we choose the daily count of bike rentals, observed in the years 2011-2012 (731 observations). They are displayed in the left panel of \autoref{fig:counts}.

There is an obvious seasonal pattern and a linear trend reflecting a growth of the bike share system (visualized by the dashed line which is the least square linear line). While the seasonal pattern will be handled by the covariates, we cannot account for the linear trend. Therefore we remove the linear trend by dividing each observation by the least squares estimate of the linear trend. We use the division rather than the subtraction of the trend since the trend is a measure for the overall members of the bike sharing community and we are interested in the proportion of members renting bikes. The resulting detrended response is plotted in the right panel of \autoref{fig:counts}.\\
For each day we have continuous covariates \textit{temperature} (apparent temperature in Celsius), \textit{wind speed} (in mph) and \textit{humidity} (relative in \%). Additionally, there is the discrete variable \textit{weather situation} giving information about the overall weather with values 1 (clear to partly cloudy), 2 (misty and cloudy) and 3 (rain, snow, thunderstorm). Further, we have information about the \textit{season} (spring, summer, fall and winter), \textit{month} and \textit{weekday} of the observed day and an indicator whether the day is a \textit{working day}.

We applied all quantile regression methods discussed in this paper to the bike sharing data set for the quantile levels 0.1, 0.5 and 0.9 and use 10-fold cross-validation to evaluate their out-of-sample performance. \autoref{tab:tick-loss} displays the corresponding averaged cross-validated tick-losses \cite[see e.g.][]{komunjer2013quantile}, given by $\frac{1}{731}\sum_{i=1}^{731}\rho_{\alpha}(y^{(i)}-\hat q_{\alpha}^{(i)})$, where $\rho_{\alpha}(y)=y(\alpha-\mathbbm{1}(y<0))$ denotes the check function, $y^{(i)}$ is the $i$-th observation of the response and $\hat q_{\alpha}^{(i)}$ is the $\alpha$-quantile prediction.  As before, the smallest losses and those which are not significantly larger than the smallest losses are printed in bold. Again, a Student's t test at 5\% level was used to test whether larger values are significantly larger than the smallest value in a row.

\begin{table}[htbp]
	\centering
		\renewcommand{\arraystretch}{1.1}
		\captionsetup{font=normalsize}
	\begin{tabular}{|l|r|r|r|r|r|}
		\hline
		$\alpha$ & PDVQR & NPDVQR & LQR & BAQR & NPQR \\ 
		\hline
		0.1 & \textbf{0.039} & \textbf{0.035} & 0.041 & \textbf{0.035} & 0.090 \\ 
		0.5 & 0.082 & \textbf{0.069} & 0.078 & \textbf{0.064} & 0.250 \\ 
		0.9 & 0.042 & \textbf{0.032} & 0.036 & \textbf{0.032} & 0.295 \\ 
		\hline
	\end{tabular}
	\caption{Averaged in-sample tick-losses of the different quantile regression methods applied to the bike sharing data.}
	\label{tab:tick-loss}
\end{table}
NPDVQR and BAQR produce the best results, significantly beating LQR and NPQR. Between the two new D-vine copula based quantile regression methods introduced in this paper, the nonparametric one significantly outperforms the parametric one for $\alpha=0.5$ and $\alpha=0.9$. The reason is that most of the covariates enter the models in a non-monotone fashion, as we will see. The ranking of the covariates by the nonparametric sequential selection algorithm is: temperature --- humidity --- wind speed --- month --- weather situation --- weekday --- working day --- season.

    \begin{figure}[!b]
    	\centering
    	\includegraphics[trim=0.7cm 0cm 2.2cm 0cm,clip,width=0.24\textwidth]{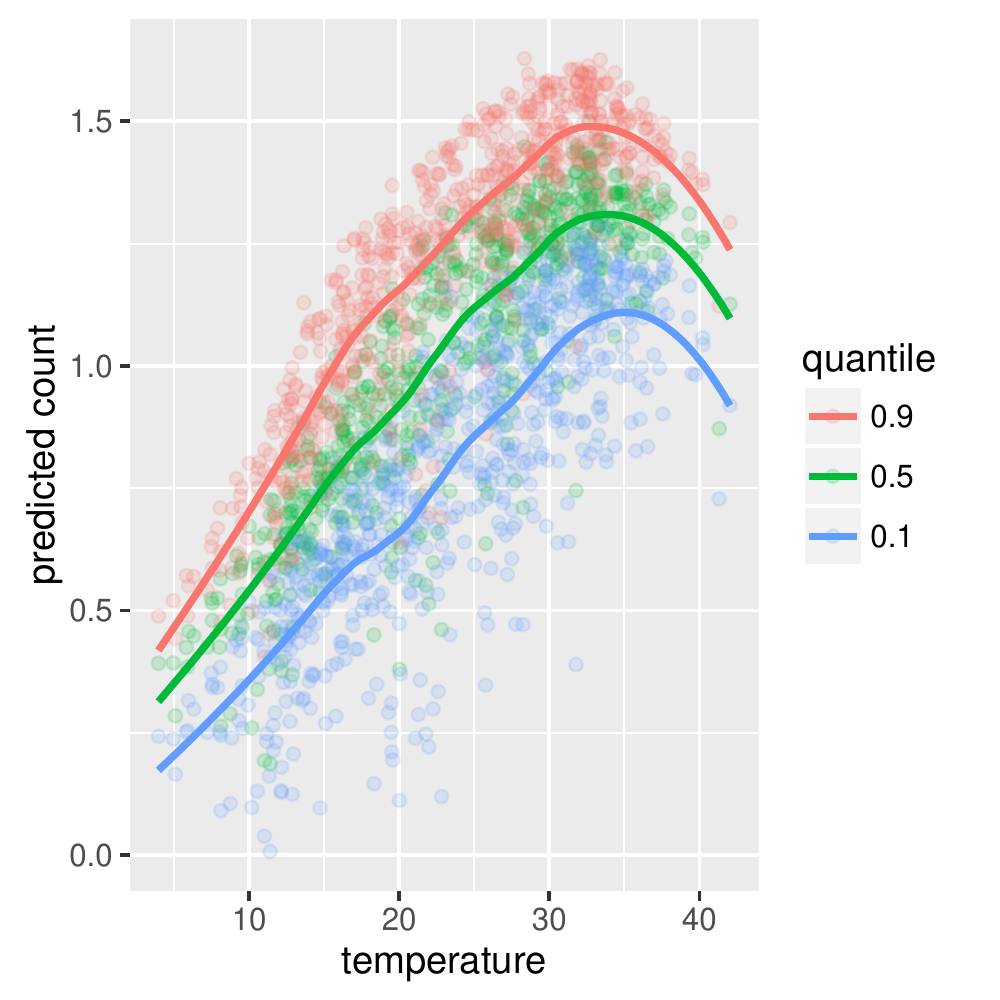}       
    	\includegraphics[trim=0.7cm 0cm 2.2cm 0cm,clip,width=0.24\textwidth]{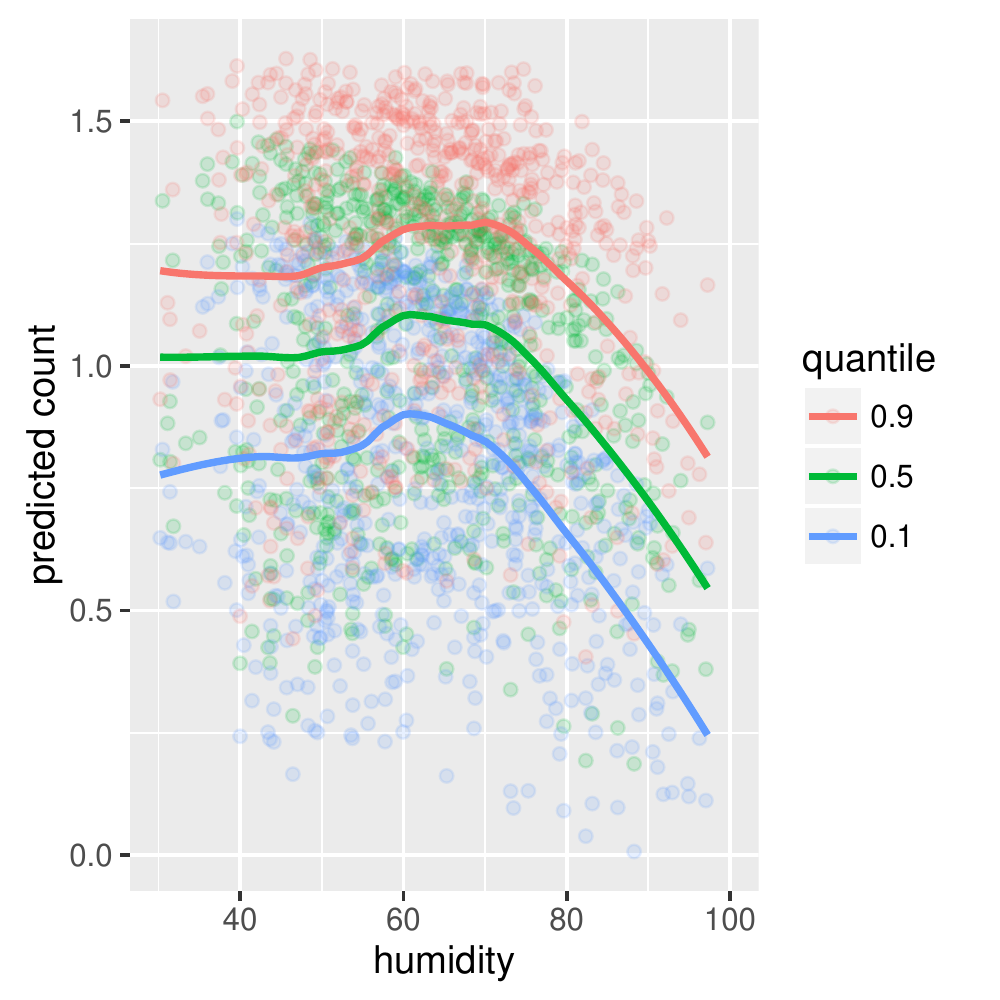}
    	\includegraphics[trim=0.7cm 0cm 2.2cm 0cm,clip,width=0.24\textwidth]{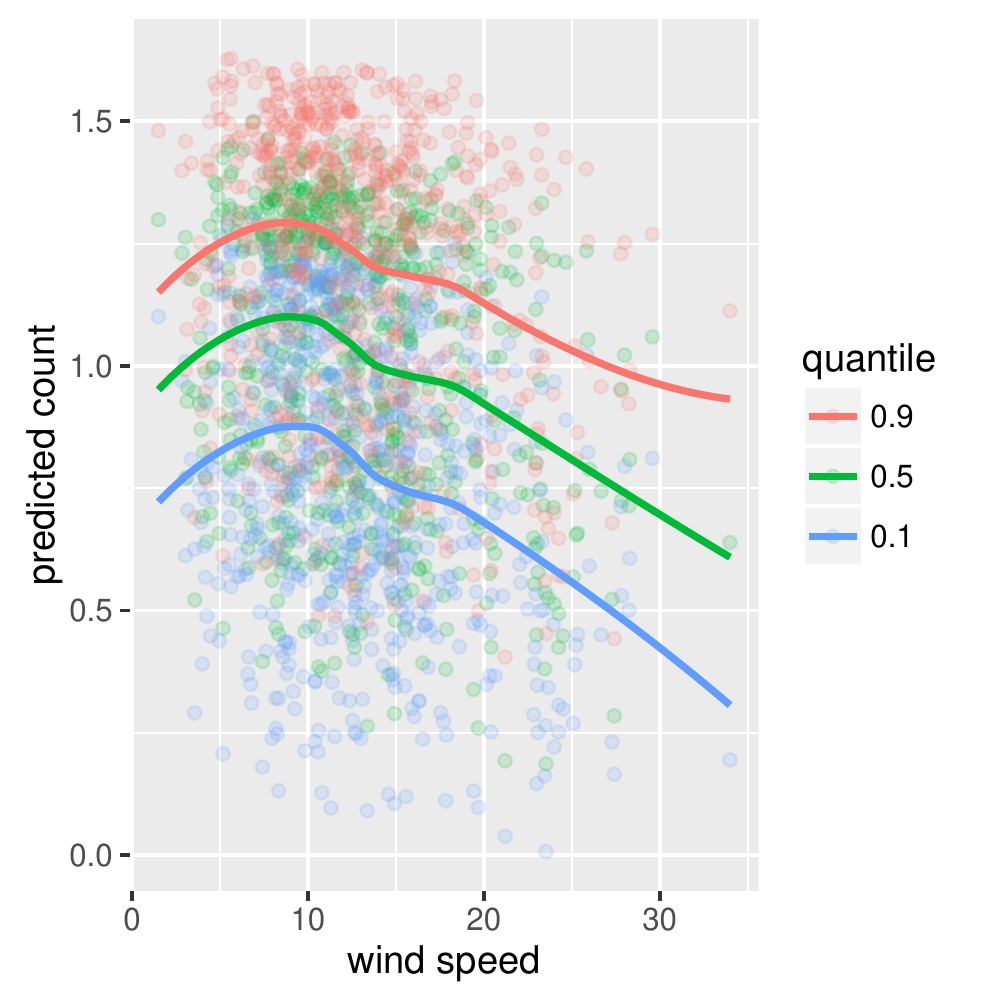}
    	\includegraphics[trim=0.7cm 0cm 2.2cm 0cm,clip,width=0.24\textwidth]{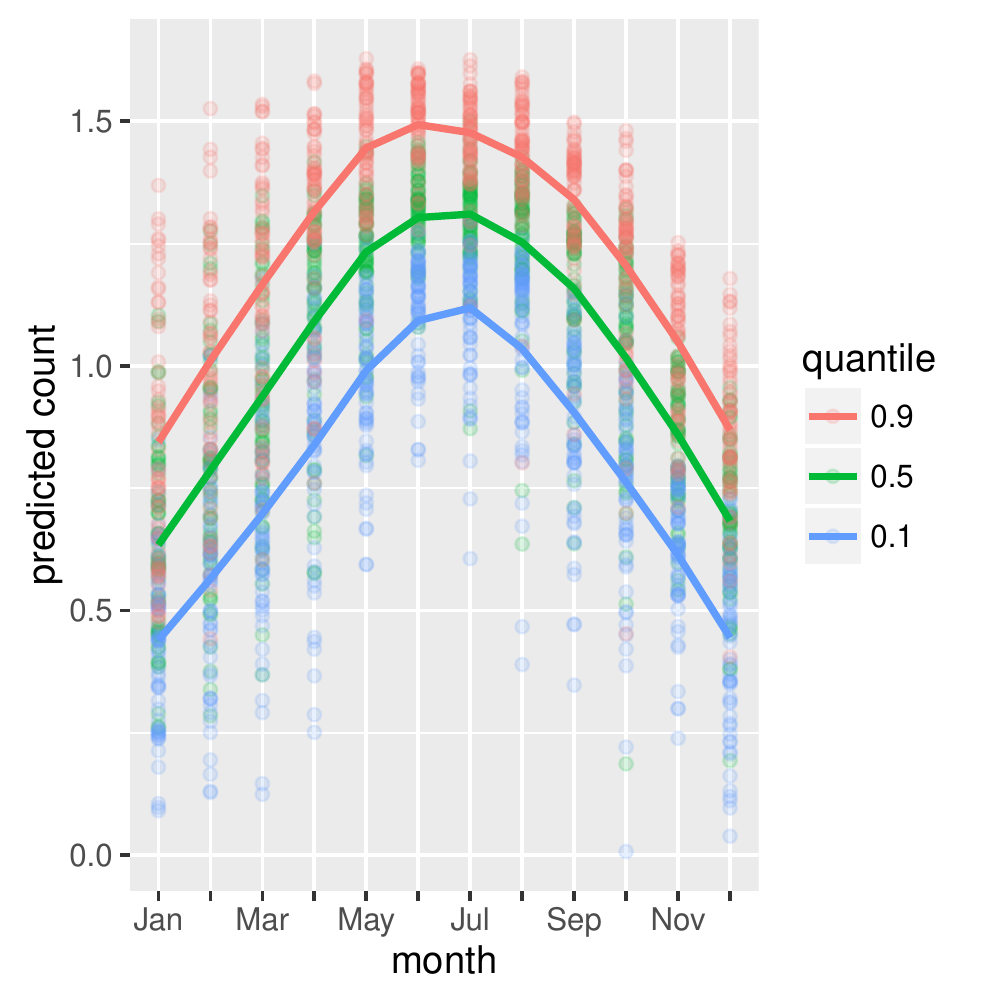}
    	\includegraphics[trim=0.7cm 0cm 2.2cm 0cm,clip,width=0.24\textwidth]{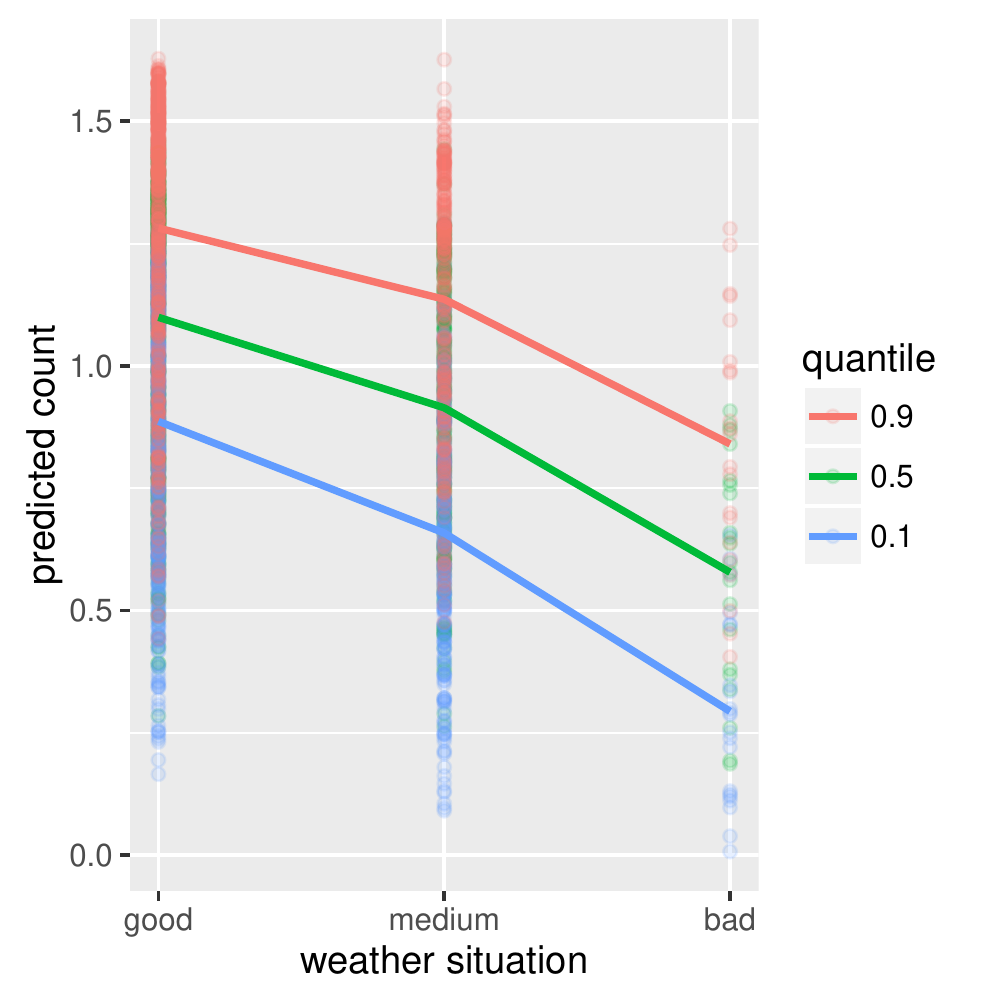}
    	\includegraphics[trim=0.7cm 0cm 2.2cm 0cm,clip,width=0.24\textwidth]{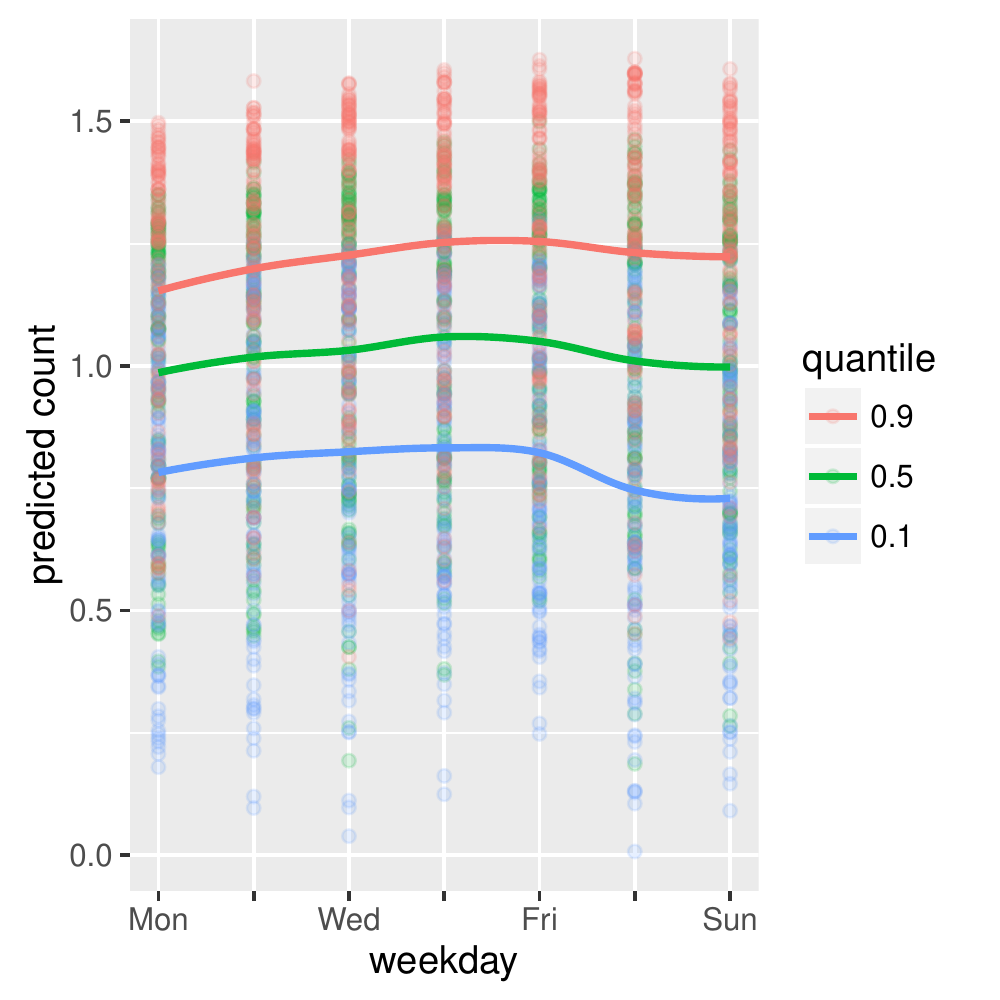}
    	\includegraphics[trim=0.7cm 0cm 2.2cm 0cm,clip,width=0.24\textwidth]{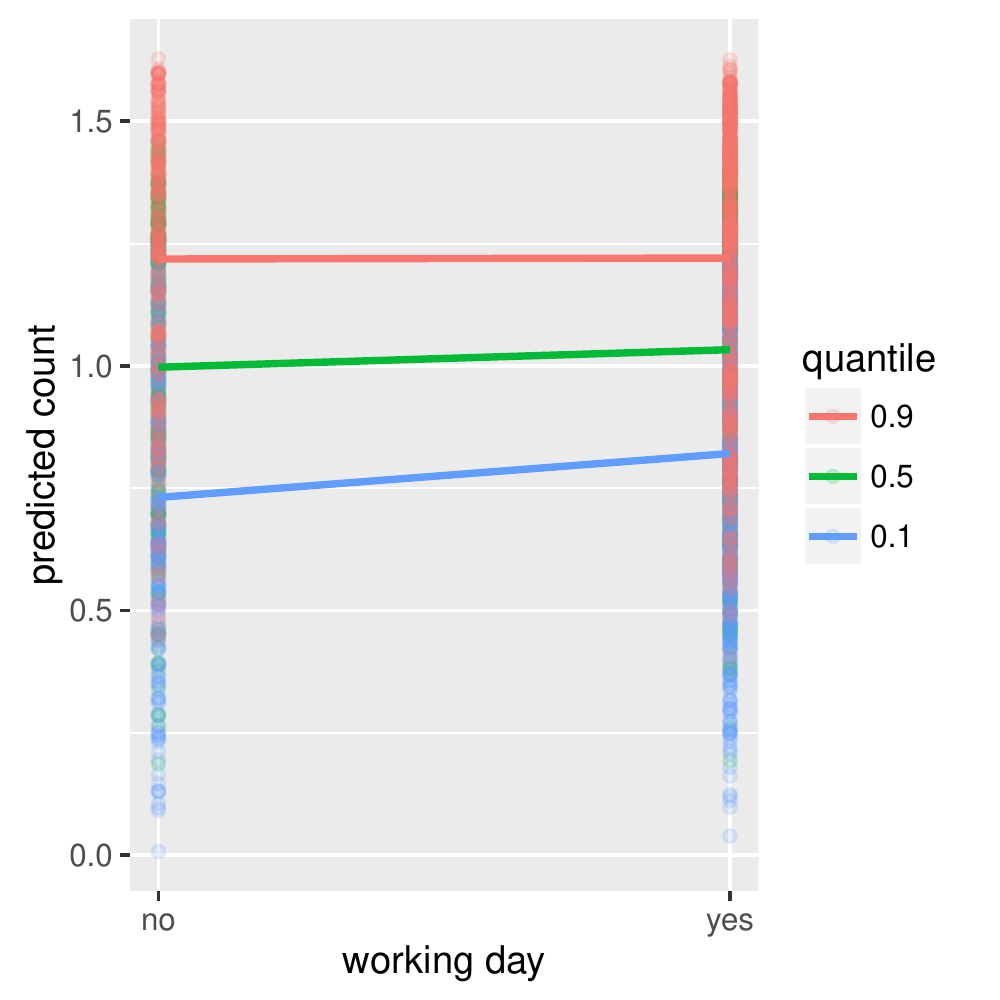}
    	\includegraphics[trim=0.7cm 0cm 2.2cm 0cm,clip,width=0.24\textwidth]{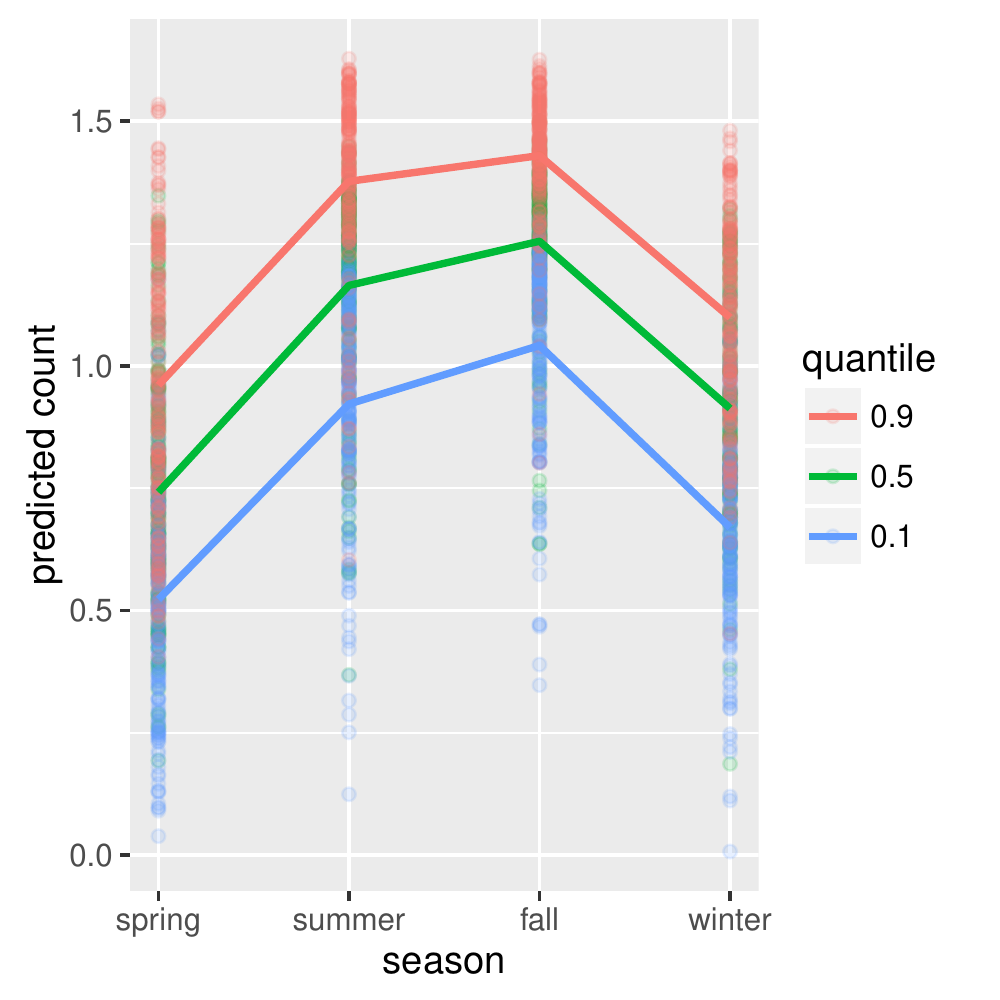}
    	\caption{Influence of the different covariates on the response bike rentals using NPDVQR.}
    	\label{fig:covEff}
    \end{figure}
    
In \autoref{fig:covEff} the influence of each of the covariates in the nonparametric D-vine quantile regression model is visualized. To be precise, for a covariate $X_j$ we calculate for all quantile levels $\alpha$ of interest $\hat{q}^{(i)}_{\alpha}=\hat F^{-1}_{Y|\Xb}(\alpha|\Xb=\xb^{(i)})$, $i=1,\ldots,731$, plot it against $x_{ij}$ and add a smooth curve through the point cloud (fitted by \texttt{loess}). \autoref{fig:covEff} shows this for the quantile levels $0.1$ (lower line), $0.5$ (middle line) and $0.9$ (upper line).

Higher temperatures generally go along with more bike rentals, until it gets too warm. For temperatures higher than 32 degrees Celsius, each additional degree causes a decline in bike rentals. Similar observations can be made for humidity. Bike rentals increase up to a relative humidity of around 60\% and decrease afterwards. Wind speed also has a strong influence with fewer bike rentals on windy days. It is not surprising that the warm summer months encourage many citizens to rent bikes while in the cold winter rentals decrease on average by approximately 60\%. The inclination to borrow bikes seems to grow during the week. On the weekend however, especially the 10\% quantile drops considerably, which may be explained by many people leaving the city to visit their families or doing leisure activities on weekends. This is also supported by the influence of variable working day, with a few more rentals on working days. The variables weather situation and season support the thesis that more people tend to rent bicycles when the weather is good.

\begin{figure}[htbp]
    	\centering
    	\includegraphics[trim=0.7cm 0cm 2.2cm 0cm,clip,width=0.19\textwidth]{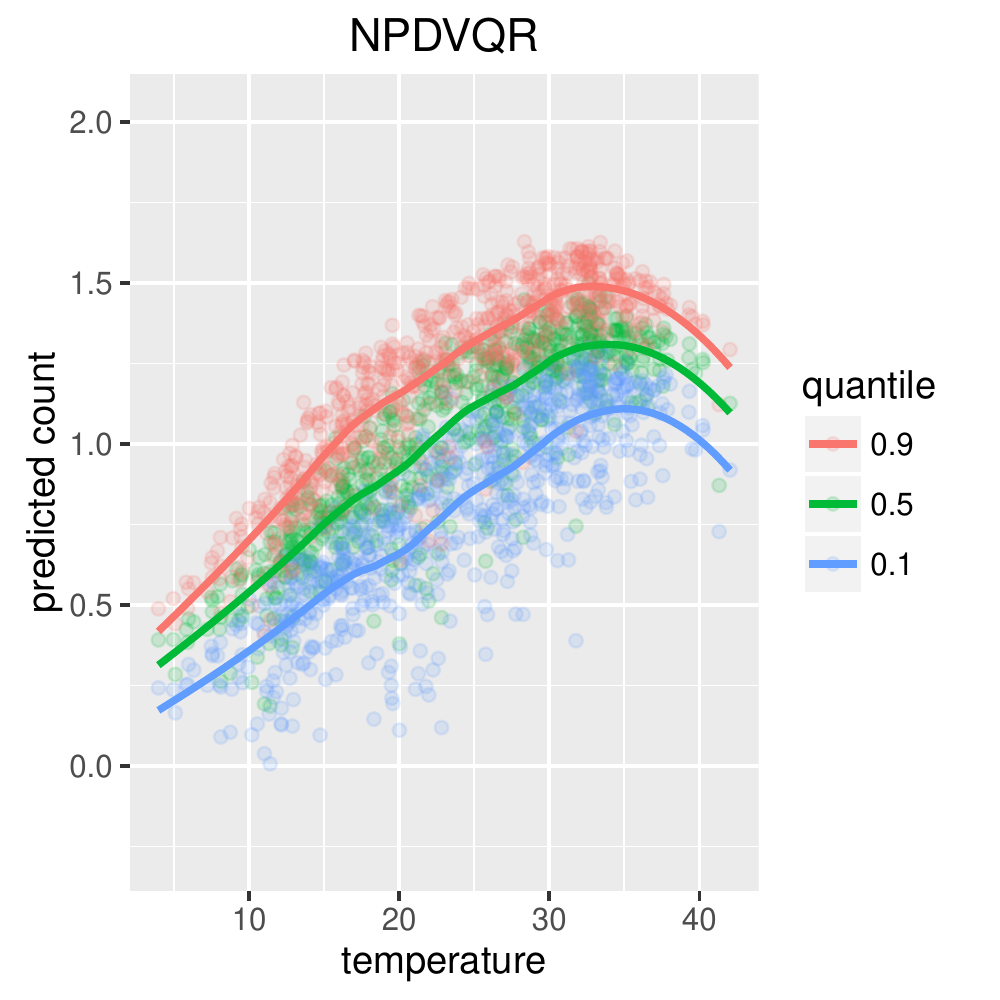}       
    	\includegraphics[trim=0.7cm 0cm 2.2cm 0cm,clip,width=0.19\textwidth]{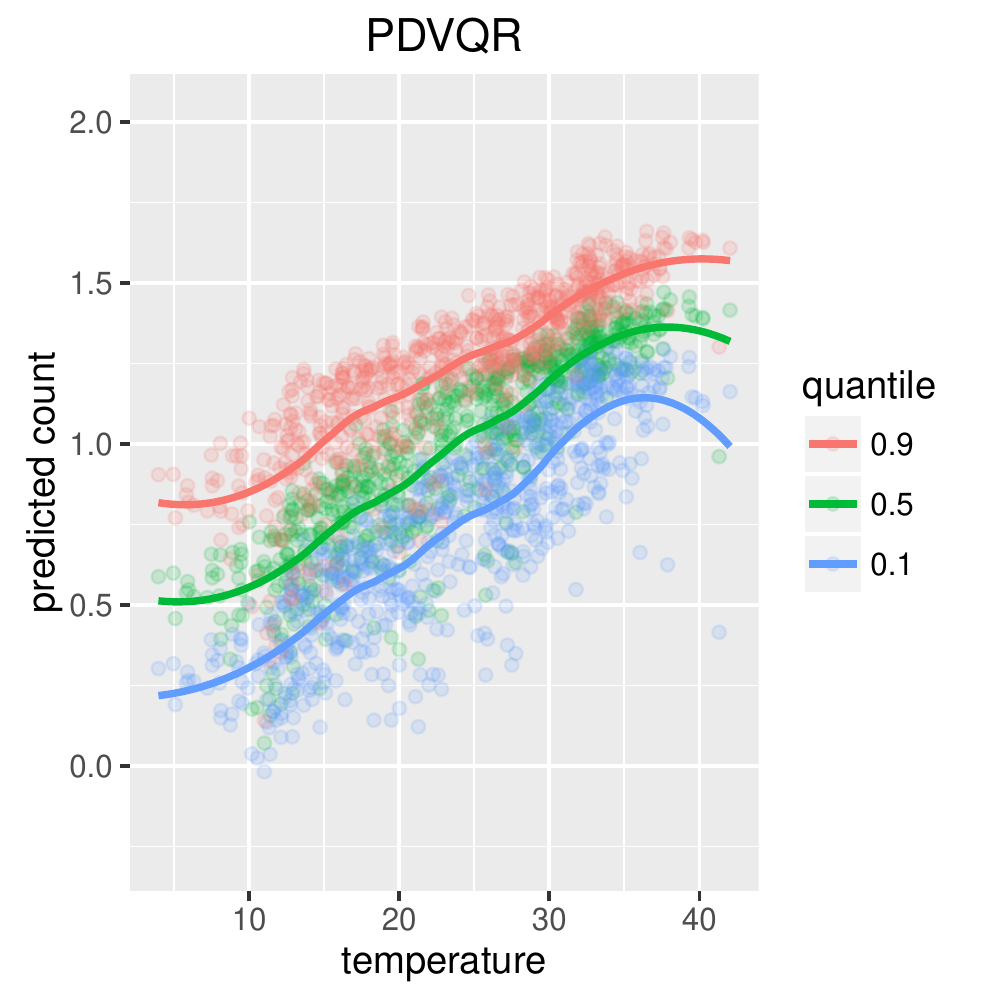}
    	\includegraphics[trim=0.7cm 0cm 2.2cm 0cm,clip,width=0.19\textwidth]{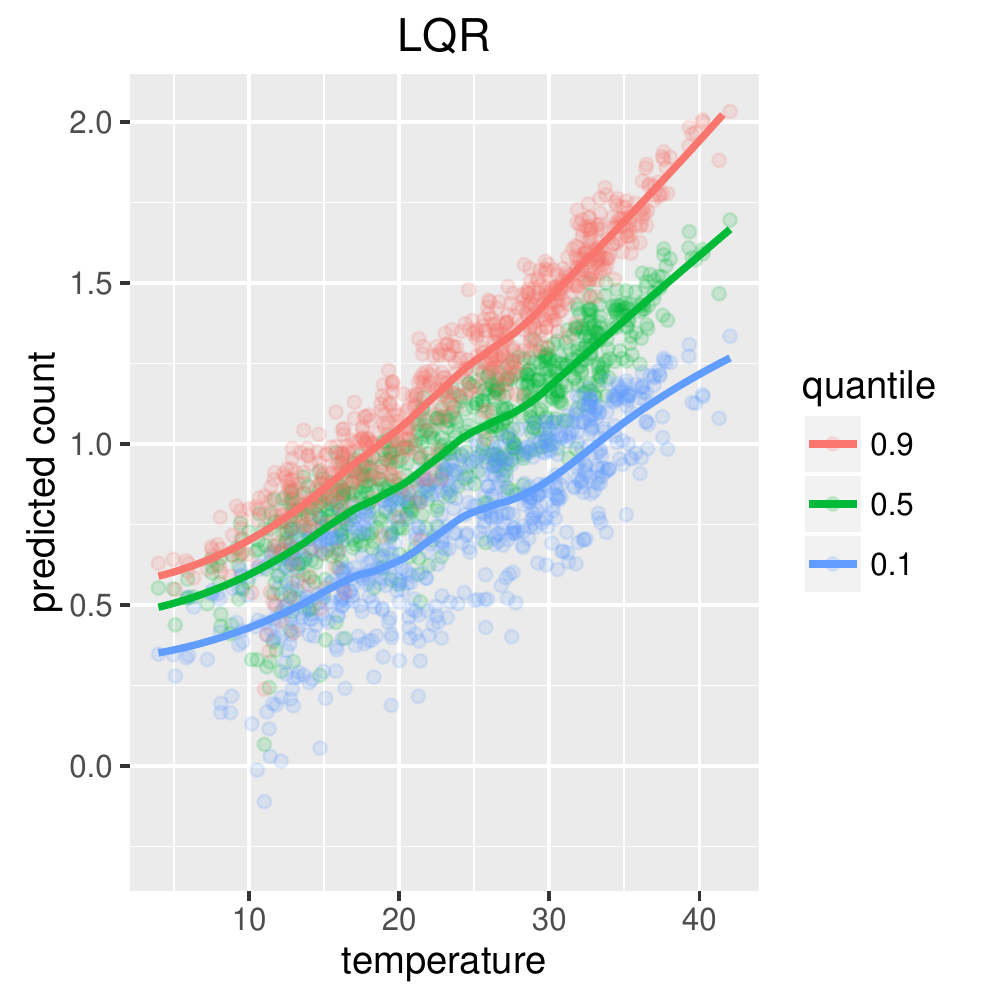}
    	\includegraphics[trim=0.7cm 0cm 2.2cm 0cm,clip,width=0.19\textwidth]{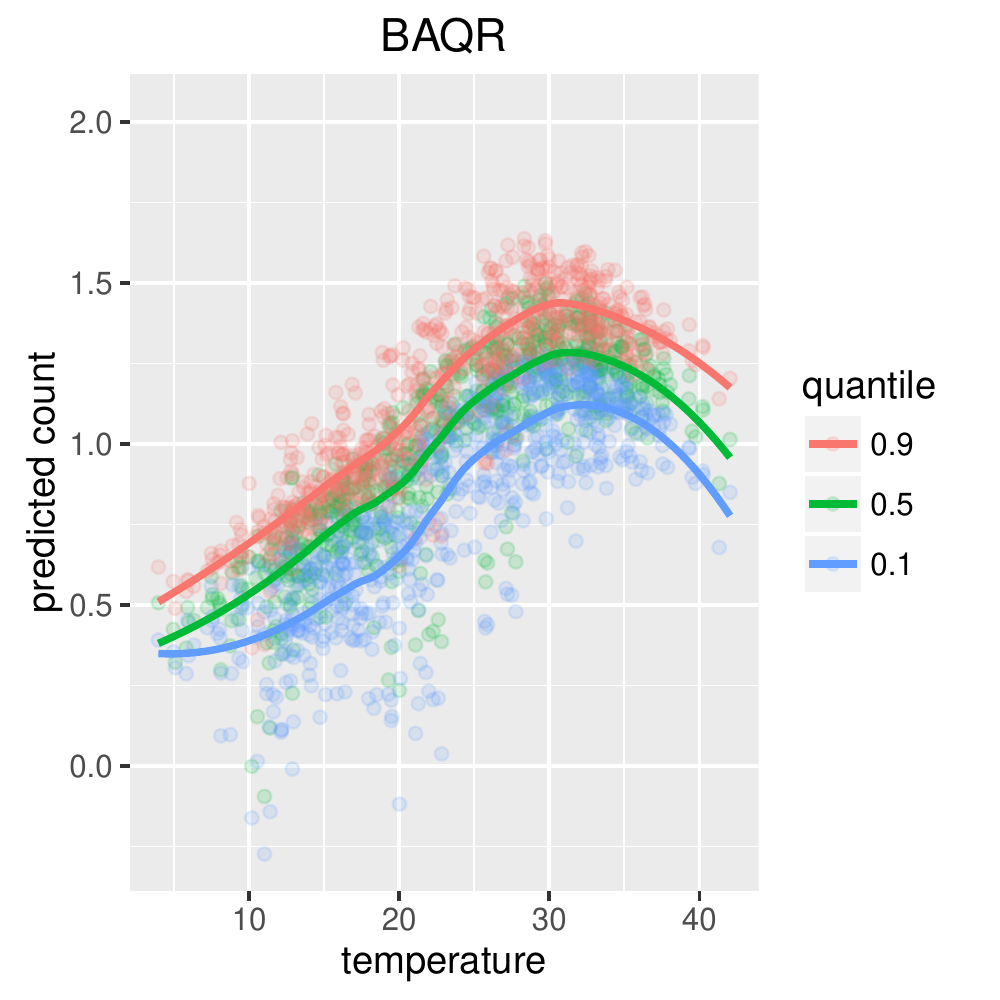}
    	\includegraphics[trim=0.7cm 0cm 2.2cm 0cm,clip,width=0.19\textwidth]{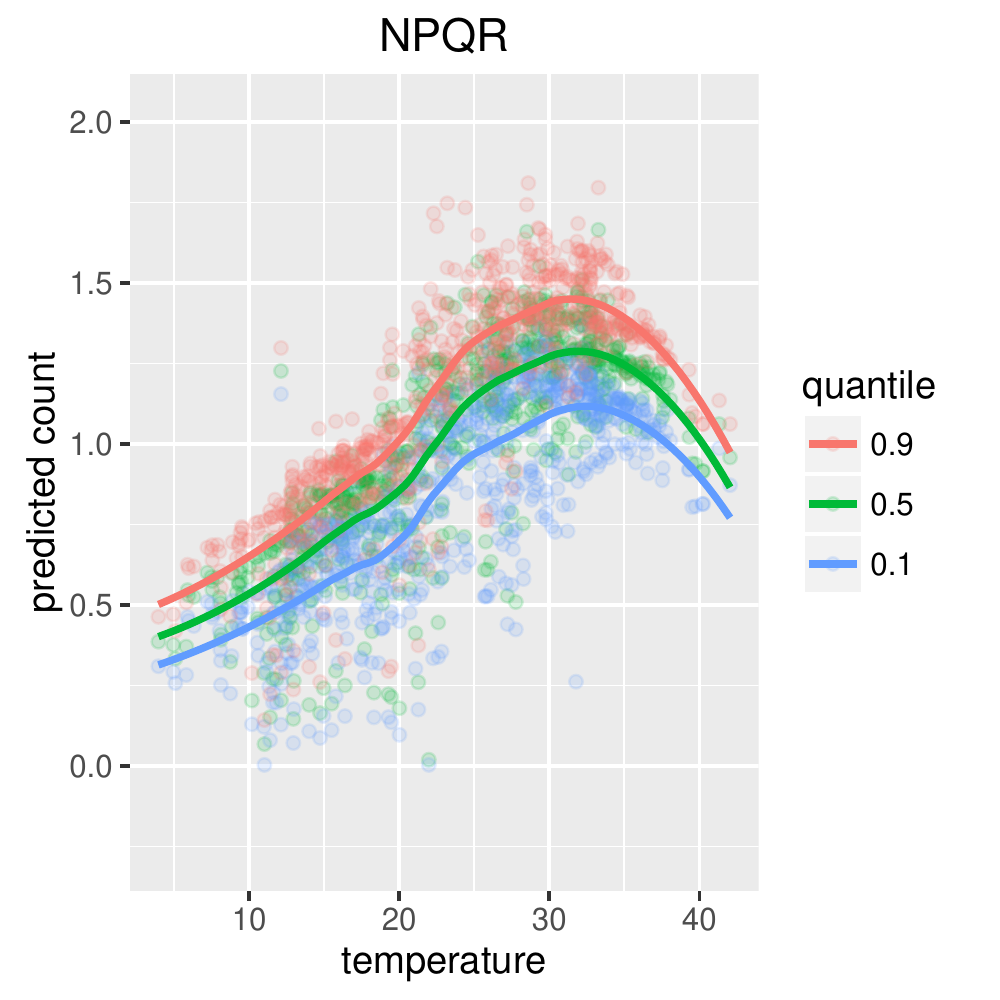}
    	\caption{Influence of temperature on bike rentals for different quantile regression methods.}
    	\label{fig:Comparison}
    \end{figure}

To investigate the differences between predictions of the various methods, we shall look more closely at the temperature variable. \autoref{fig:Comparison} shows the effect of temperature on the predicted bike rentals using NPDVQR, PDVQR, LQR, BAQR and NPQR (from left to right).

We see that the parametric D-vine as well as linear quantile regression are not really able to model the decline in rentals for very hot temperatures.\\
%Further, the purely nonparametric and the boosted additive quantile estimations seem to have the issue of over-fitting with declining rentals for temperatures going from 0 to 10 degrees Celsius.\\
Apart from assessing the influence of covariates on the response, quantile regression can also be used to predict quantiles of the response in different scenarios. Suppose we know tomorrow is going to be a warm August Saturday with medium humidity and low wind-speed. Then, using our nonparametric D-vine copula based quantile regression model, we would predict a median of 8872 bikes to be rented with 10\%- and 90\% quantiles 7431 and 10485, respectively. In contrast, for a cold December Monday with heavy snow and high wind-speed the three predicted quantiles would be 22, 674 and 1152. As an operator of such a bike sharing system we could thus adapt our supply of rental bikes to the predicted demand.

\section{Conclusion and outlook}
\label{cha6}
Two new methods to predict conditional quantiles in a mixed discrete-continuous setting are proposed. They are based on a D-vine copula model that is estimated either parametrically or nonparametrically. The simulation study shows that the non-parametric D-vine quantile regression provides fast and accurate predictions for non-linear relationships between the quantile and the covariates.
The parametric approach is often less accurate.
This is due to the fact that non-linear relationships imply non-monotonic effects of some covariates on the response,
which cannot be adequately modeled by most of the popular parametric pair-copula families. This shortfall could be overcome by using parametric families  that allow for non-monotonic dependence patterns. Developing such models will be a promising path for future research.

% Ich befürchte, das ist gar nicht möglich.
%A second field of future research could be the derivation of analytical expressions for the inverses of the $ \widetilde{\text{h}} $-functions.Then, the numerical inversion of the conditional distribution functions currently used in the parametric D-vine quantile regression could be replaced by expressing the conditional quantiles by nested inverted h- and $ \widetilde{\text{h}} $-functions, as it is done for the continuous case in \cite{kraus2017d}.

\section*{Acknowledgment}\label{sec:acknowledgment}
The third and fourth authors are supported by the German Research Foundation (DFG grants CZ 86/5-1 and CZ 86/4-1). Numerical calculations were performed on a Linux cluster supported by DFG grant INST 95/919-1 FUGG. 

\bibliographystyle{apalike}
\bibliography{References}

\end{document}